\documentclass[letterpaper, 11pt]{article}
\usepackage{graphicx}
\usepackage{amsmath}
\usepackage{amsfonts}
\usepackage{hyperref}
\usepackage{amsthm}
 \usepackage{amscd}
\usepackage{mathrsfs}
\usepackage{caption}
\usepackage{subcaption}
\usepackage{float}
\usepackage[top=1in,left=1in,right=1in,bottom=1in]{geometry}

\counterwithin{figure}{section}
\hypersetup{
	colorlinks = true,
	linkcolor  = red,
	citecolor  = blue,
	filecolor  = cyan,
	urlcolor   = magenta,}

\newcommand{\ds}{\displaystyle}

\numberwithin{equation}{section}

\title{Inverse scattering problem for the third-order
equation on the line}
\author{Tuncay Aktosun and Ivan Toledo\\
Department of Mathematics\\
University of Texas at Arlington\\
Arlington, TX 76019-0408, USA\\
\\
Mehmet Unlu\\
Department of Mathematics\\
Recep Tayyip Erdogan University\\
53100 Rize, Turkey
\\
\\
\\
\textit{Dedicated to the memory of Pierre C. Sabatier, a pioneer in inverse problems}
}

\date{}

\begin{document}

\maketitle

\begin{abstract}
We consider the third-order linear differential equation
$$\displaystyle\frac{d^3\psi}{dx^3}+Q(x)\,\displaystyle\frac{d\psi}{dx}+P(x)\,\psi=k^3\,\psi,\qquad x\in\mathbb R,$$
where the complex-valued potentials $Q$ and $P$ are assumed to belong to the Schwartz class. We describe the basic solutions, the scattering coefficients, and the bound-state
information, and we introduce the dependency constants and the normalization constants at the bound states.
When the secondary reflection coefficients
are zero, we provide a method to solve the corresponding inverse scattering problem, where
the goal is to recover the two potentials
$Q$ and $P$ from the scattering data set consisting of the transmission and primary reflection
coefficients and the bound-state information. We formulate the corresponding inverse scattering problem as a 
Riemann--Hilbert problem on the complex $k$-plane and describe how the potentials are recovered from the solution to the
Riemann--Hilbert problem. In the absence of bound states, we introduce a linear integral equation, which is the analog of the Marchenko integral equation used in the 
inverse scattering theory for the full-line Schr\"odinger equation. We describe the recovery of the two potentials from the solution to
the aforementioned linear integral equation.
\end{abstract}



\newpage

\section{Introduction}
\label{section1} 
In this paper, we analyze the direct and inverse scattering problems for the third-order equation
\begin{equation}
\label{1.1}
\psi'''+Q(x)\,\psi'+P(x)\,\psi=k^3\,\psi, \qquad x\in\mathbb R,
\end{equation}
where $x$ is the independent variable taking values on the real axis $\mathbb R,$
the prime denotes the $x$-derivative, $k^3$ is the spectral parameter, $\psi$ is the wavefunction, and $Q$ and $P$
are the complex-valued potentials satisfying certain restrictions. For simplicity, we assume that $Q$ and $P$ belong to
the Schwartz class $S(\mathbb R)$ even though our results hold under weaker assumptions on those two potentials.
We recall that the Schwartz class consists of infinitely differentiable functions $\theta(x)$ in such a way that, for
any pair of nonnegative integers $j$ and $l,$ the quantity $x^j d^l\theta(x)/dx^l$ vanishes as $x\to\pm\infty.$

The direct scattering problem for \eqref{1.1} consists of the determination of the scattering coefficients and
the bound-state information when the potentials $Q$ and $P$ are known. On the other hand, the inverse scattering problem
for \eqref{1.1} deals with the determination of the potentials $Q$ and $P$ when the scattering coefficients and the bound-state
information are known. The bound-state information usually consists of the bound-states poles of a transmission coefficient
and the specification of a dependency constant or a normalization constant at each bound state.

The physical importance of the analysis of the inverse scattering problem for \eqref{1.1} comes from the fact that the solution to that inverse problem, via the inverse scattering transform method
\cite{GGKM1967,L1968}, is used to determine solutions to various integrable nonlinear partial differential equations arising in 
fluid dynamics. For example, the fourth-order nonlinear partial differential equation \cite{BZ2002,B1872,CL2024} introduced by the French mathematician and physicist Boussinesq
is used to describe the propagation of surface water waves when those waves have relatively long wavelength and propagate in a weakly nonlinear medium. The solution to the inverse problem for \eqref{1.1} is also relevant in solving the fifth-order nonlinear partial differential equations known as the Sawada--Kotera equation \cite{SK1974} and Kaup--Kupershmidt equation \cite{K1984}. Like the third-order nonlinear partial differential equation known as the Korteweg-de Vries equation \cite{KdV1895}, the Sawada--Kotera equation and the Kaup--Kupershmidt equation are used to describe the propagation of surface water waves in long, shallow, narrow canals when those waves have steeper wavefronts and shorter wavelengths. 

The mathematical significance of our paper is that it introduces
a linear integral equation, which is the analog of the Marchenko integral equation \cite{CS1989,DT1979,F1967,L1987,M2011} used in the inverse scattering theory for 
the Schr\"odinger equation, to solve
the inverse scattering problem for \eqref{1.1}.  
It also establishes a properly posed Riemann-Hilbert problem to solve
the inverse scattering problem for \eqref{1.1}.  Thus, our paper enables the use of the inverse scattering transform method to solve
integrable evolution equations associated with the third-order linear operator for \eqref{1.1}. Hirota's bilinear method \cite{H1973,H1989,P2000,P2001}
has been the only viable
method to obtain soliton solutions to such integrable evolution equation.
However, Hirota's bilinear method is an ad hoc algebraic procedure without any physical motivation.
Parker stated in his 2000 paper \cite{P2000} that “it seems to us that obtaining the multisoliton
solutions to the Kaup–Kupershmidt equation via IST (Inverse Scattering Transform), or one of
its variants, remains a worthwhile goal.” The same observation is true also for other integrable evolution equations such as
the Sawada--Kotera equation and the Boussinesq equation.

Contrary to the second-order linear differential equations such as the Schr\"odinger equation, the theory of inverse scattering problem for the third-order equation \eqref{1.1} is not well developed. Kaup \cite{K1980} started the study
 of the direct and inverse scattering problems for \eqref{1.1} with the aim of finding solutions to the Sawada--Kotera equation and the Kaup--Kupershmidt equation. He always desired \cite{K2002}
 to find the analog of the Marchenko integral equation for \eqref{1.1} so that the solution to that linear integral equation
  would yield
 solutions to the Sawada--Kotera equation and the Kaup--Kupershmidt equation. Kaup failed \cite{H1989,K2002,P2001} in his efforts
 to obtain such a linear integral equation. In fact, Kaup 
 was not even able to formulate \cite{K1980} a proper scattering matrix to analyze the direct scattering problem for \eqref{1.1}. Consequently, he was not able to formulate a proper Riemann--Hilbert problem related to the inverse scattering theory for \eqref{1.1}. We refer to a Riemann--Hilbert problem as properly posed when that Riemann--Hilbert problem is formulated along a full line separating the complex plane into two half planes, as in the formulation used in the inverse scattering theory for the Schr\"odinger equation. A proper formulation of the Riemann--Hilbert problem allows the use of a Fourier transformation along the aforementioned full line separating the two complex half planes. With the help of the Fourier transformation, a
 properly posed Riemann--Hilbert problem allows the formulation of a linear integral equation, which is the analog of the Marchenko integral equation 
 for the Schr\"odinger equation. Deift, Tomei, and Trubowitz studied \cite{DTT1982} the direct and inverse scattering problems for a special case of \eqref{1.1}, when the corresponding third-order linear differential operator is selfadjoint. They introduced \cite{DTT1982} a proper Riemann--Hilbert problem in the complex plane to solve the corresponding inverse scattering problem. Their physical motivation was to analyze the solution to a modified version \cite{DTT1982} of the Boussinesq equation. However, in order to formulate their Riemann--Hilbert problem properly, Deift, Tomei, and
 Trubowitz made \cite{DTT1982} the severe assumption that the two transmission coefficients are identically equal to 1 for all $k$-values in the complex plane.
They also made the assumption that the two secondary reflection coefficients are identically zero for all $k$-values. Their restriction on the secondary reflection coefficients is not as severe as their restriction on the transmission coefficients, and in fact the former restriction helps avoid solutions to the modified Boussinesq equation that blow up at a finite time. The severe restriction on the transmission coefficients prevented Deift, Tomei, and Trubowitz from obtaining explicit solution to the modified Boussinesq equation. The reason for this is that explicit solutions to integrable nonlinear partial differential equations are usually obtained when all the reflection coefficients are zero and by using the bound-state poles of the transmission coefficients. Inspired by the important paper \cite{DTT1982} by Deift, Tomei, and Trubowitz, the second author of the present paper in his recent Ph.D. thesis \cite{T2024} formulated
a proper Riemann--Hilbert problem for the inverse scattering theory for \eqref{1.1} by using only the assumption that the two secondary reflection coefficients are zero but without using the assumption that the transmission coefficients are identically equal to 1 for all $k$-values.

The most relevant papers in the literature for our work is the 1980 paper \cite{K1980} by Kaup, the 1982 paper \cite{DTT1982} by Deift, Tomei, and Trubowitz, the 2024 Ph.D. thesis \cite{T2024} by Toledo, and the recent paper \cite{ACTU2025} where soliton solutions to some integrable nonlinear evolution equations are obtained by solving our properly formulated
Riemann--Hilbert problem when all the reflection coefficients are zero. We also mention the references
\cite{B1985,BC1984,BC1987,C1980,K2002} where the Riemann--Hilbert problem is formulated on a set of two intersecting full lines on the complex plane. However, such a formulation
does not allow the use of scattering coefficients for \eqref{1.1} as input to solve the corresponding inverse scattering problem.

Our paper is organized as follows. In Section~\ref{section2}, we introduce the four basic solutions to \eqref{1.1} and the six scattering coefficients associated with \eqref{1.1}, and we provide the relevant properties of those solutions and scattering coefficients. In Section~\ref{section3}, we present a brief description of the bound states for \eqref{1.1}, and we introduce the dependency constants and the normalization constants for the bound states. In Section~\ref{section4}, we consider the important special case where the secondary reflection coefficients are zero but without the severe restriction on the transmission coefficients used \cite{DTT1982} by Deift, Tomei, and Trubowitz. We formulate the relevant proper Riemann--Hilbert problem, and we describe the recovery of the potentials $Q$ and $P$ from the solution to that Riemann--Hilbert problem. In Section~\ref{section5}, we derive the linear integral equation associated with \eqref{1.1}, which is the analog of the Marchenko equation 
used in the inverse scattering theory for the Schr\"odinger equation. For simplicity, we derive our linear integral equation in the absence of bound states. We also describe the recovery of the potentials $Q$ and $P$ from the solution to the aforementioned integral equation.

\section{The basic solutions and scattering coefficients}
\label{section2}

We recall that we assume that the potentials $Q$ and $P$ appearing in \eqref{1.1} belong to the Schwartz class. 
In this section we present the four basic solutions to \eqref{1.1}, denoted by $f(k,x),$ $g(k,x),$ $m(k,x),$ $n(k,x),$ respectively, and the six
scattering coefficients denoted by $T_{\text{\rm{l}}}(k),$ $L(k),$ $M(k),$ $T_{\text{\rm{r}}}(k),$ $R(k),$ $N(k),$ respectively. 
We provide the large spacial and spectral asymptotics of the four basic solutions, and we present some relevant properties of the scattering coefficients.
For further details and proofs, we refer the reader to \cite{ACTU2025,T2024}.

For each fixed $k\in\mathbb C,$ we are interested in identifying a set of three fundamental solutions to \eqref{1.1}. Since \eqref{1.1} is a linear
homogeneous differential equation, we can use three fundamental solutions to \eqref{1.1} at a $k$-value to determine the general solution.
We remark that if $\psi(k,x)$
is a solution to \eqref{1.1}, then $\psi(zk,x)$ and $\psi(z^2k,x)$ are also solutions, where $z$ is the special complex constant $e^{2\pi i/3},$ or equivalently 
\begin{equation}\label{2.1}
z:=-\ds\frac{1}{2}+i\,\ds\frac{\sqrt{3}}{2}.
\end{equation}
This is because the parameter $k$ appears as $k^3$ in \eqref{1.1} and we have $(zk)^3=k^3$ and $(z^2k)^3=k^3.$
The $k$-domain of $\psi(zk,x)$ is obtained by rotating the $k$-domain of $\psi(k,x)$
around the origin of the complex $k$-plane by $2\pi/3$ radians clockwise. Similarly, the $k$-domain of $\psi(z^2k,x)$ is obtained by rotating the
$k$-domain of $\psi(k,x)$ by $4\pi/3$ radians clockwise. This observation is useful to construct a fundamental set of three solutions to \eqref{1.1}
at each $k$-value.

\begin{figure}[!ht]
     \centering
         \includegraphics[width=2.1in]{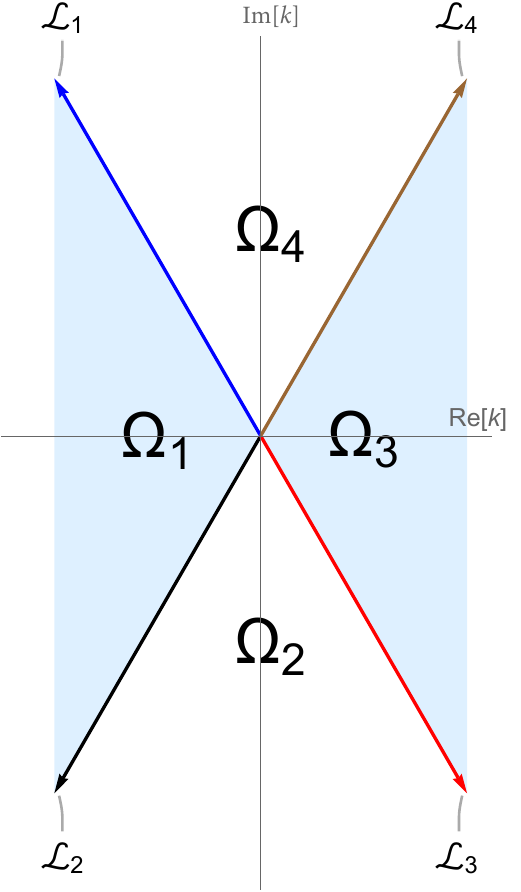}      \hskip .3in
         \includegraphics[width=2.25in]{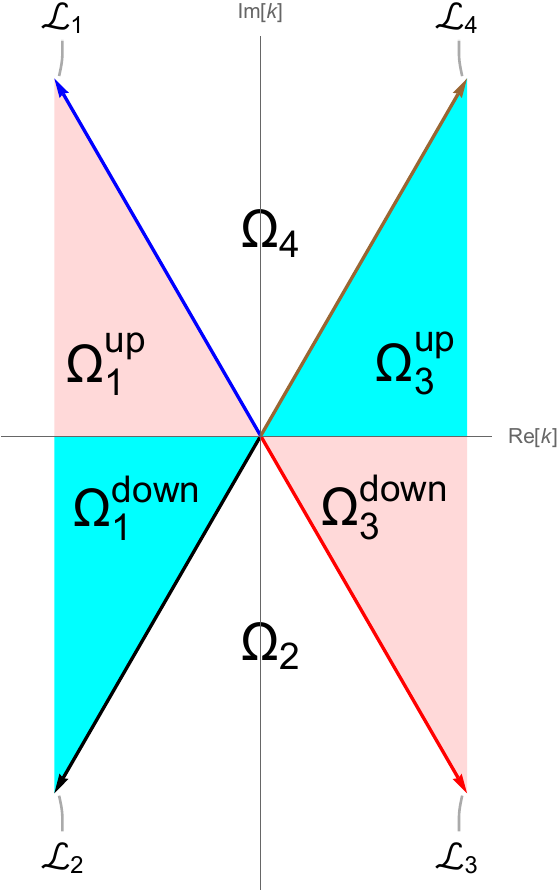} 
\caption{The complex $k$-plane is divided into the 
four sectors
 $\Omega_1,$ 
$\Omega_2,$ $\Omega_3,$ and $\Omega_4$ as shown on the left plot,
with the directed half lines $\mathcal L_1,$ $\mathcal L_2,$ $\mathcal L_3,$ and $\mathcal L_4$
acting as the boundaries. The sectors $\Omega_1$ and $\Omega_3$ are each divided
into two subsectors by the directed half lines $-\mathbb R^-$ and $\mathbb R^+,$ respectively,
as shown on the right plot.}
\label{figure2.1}
\end{figure}

We divide the complex $k$-plane into six open sectors $\Omega_1^\text{\rm{up}},$ $\Omega_1^\text{\rm{down}},$ 
$\Omega_2,$ $\Omega_3^\text{\rm{down}},$ $\Omega_3^\text{\rm{up}},$ and $\Omega_4$ as shown in Figure~\ref{figure2.1}. These six open
sectors are parametrized, respectively, as
\begin{equation*}
\Omega_1^\text{\rm{up}}:=\left\{k\in\mathbb C: \ds\frac{2\pi}{3}<\arg[k]<\pi\right\},
\end{equation*}
\begin{equation*}
\Omega_1^\text{\rm{down}}:=\left\{k\in\mathbb C: \pi<\arg[k]<\ds\frac{4\pi}{3}\right\},
\end{equation*}
\begin{equation*}
\Omega_2:=\left\{k\in\mathbb C: -\ds\frac{2\pi}{3}<\arg[k]<-\ds\frac{\pi}{3}\right\},
\end{equation*}
\begin{equation*}
\Omega_3^\text{\rm{down}}:=\left\{k\in\mathbb C: -\ds\frac{\pi}{3}<\arg[k]<0\right\},
\end{equation*}
\begin{equation*}
\Omega_3^\text{\rm{up}}:=\left\{k\in\mathbb C: 0<\arg[k]<\ds\frac{\pi}{3}\right\},
\end{equation*}
\begin{equation*}
\Omega_4:=\left\{k\in\mathbb C: \ds\frac{\pi}{3}<\arg[k]<\ds\frac{2\pi}{3}\right\},
\end{equation*}
where we use $\arg[k]$ to denote the argument function taking values in the interval $(-2\pi/3,4\pi/3).$
We use $\overline{\Omega_1^\text{\rm{up}}},$ $\overline{\Omega_1^\text{\rm{down}}},$ $\overline{\Omega_2},$ $\overline{\Omega_3^\text{\rm{down}}},$ $\overline{\Omega_3^\text{\rm{up}}},$ 
$\overline{\Omega_4}$ to denote the closures of the respective open sets, where we recall that the closure
of an open set is obtained by including the boundary of that open set. We also define $\Omega_1$ and $\Omega_3$ as the open sectors parametrized, respectively, as
\begin{equation*}
\Omega_1:=\left\{k\in\mathbb C: \ds\frac{2\pi}{3}<\arg[k]<\ds\frac{4\pi}{3}\right\},
\end{equation*}
\begin{equation*}
\Omega_3:=\left\{k\in\mathbb C: -\ds\frac{\pi}{3}<\arg[k]<\ds\frac{\pi}{3}\right\},
\end{equation*}
and we use $\overline{\Omega_1}$ and $\overline{\Omega_3}$ to denote their closures, respectively.

The left Jost solution $f(k,x)$ is the solution to \eqref{1.1} satisfying the asymptotics as $x\to +\infty$ given by
\begin{equation}\label{2.10}
\begin{cases}
f(k,x)=e^{kx}\left[1+o(1)\right],\\
\noalign{\medskip}
f'(k,x)=k\,e^{kx}\left[1+o(1)\right],\\
\noalign{\medskip}
f''(k,x)=k^2\,e^{kx}\left[1+o(1)\right],
\end{cases}
\end{equation}
and its $k$-domain is given by $\overline{\Omega_1}.$ The left scattering coefficients $T_{\text{\rm{l}}}(k),$ $L(k),$ $M(k)$ are defined via the spacial
asymptotics of $f(k,x)$ as $x\to -\infty$ given by
\begin{equation}\label{2.11}
f(k,x)=\begin{cases}
e^{kx}\,T_{\text{\rm{l}}}(k)^{-1}[1+o(1)]+e^{zkx}L(k)\,T_{\text{\rm{l}}}(k)^{-1}[1+o(1)],\qquad k\in \mathcal L_1,\\
\noalign{\medskip}
e^{kx}\,T_{\text{\rm{l}}}(k)^{-1}[1+o(1)], \qquad k\in \Omega_1,\\
\noalign{\medskip}
e^{kx}\,T_{\text{\rm{l}}}(k)^{-1}[1+o(1)]+e^{z^2 kx}M(k)\,T_{\text{\rm{l}}}(k)^{-1}[1+o(1)], \qquad k\in \mathcal L_2,
\end{cases}
\end{equation}
where $\mathcal L_1$ and $\mathcal L_2$ are the directed half lines shown in Figure~\ref{figure2.1} corresponding to the upper and lower
boundaries of $\Omega_1,$ respectively. We note that $\mathcal L_1$ and $\mathcal L_2$ are parametrized, respectively, as
\begin{equation*}
\mathcal L_1:=\{k\in\mathbb C: k=zs \text{\rm{ for }}  s\in[0,+\infty)\},
\end{equation*}
\begin{equation*}
\mathcal L_2:=\{k\in\mathbb C: k=z^2s \text{\rm{ for }}  s\in[0,+\infty)\}.
\end{equation*}
The left transmission coefficient $T_{\text{\rm{l}}}(k)$ is defined on $\overline{\Omega_1},$ the left primary reflection coefficient $L(k)$ is defined
on $\mathcal L_1,$ and the left secondary reflection coefficient $M(k)$ is defined on $\mathcal L_2.$  

The right Jost solution $g(k,x)$ is the solution to \eqref{1.1} satisfying the spacial asymptotics as $x\to -\infty$ given by
\begin{equation}\label{2.14}
\begin{cases}
g(k,x)=e^{kx}\left[1+o(1)\right],\\
\noalign{\medskip}
g'(k,x)=k\,e^{kx}\left[1+o(1)\right],\\
\noalign{\medskip}
g''(k,x)=k^2\,e^{kx}\left[1+o(1)\right],
\end{cases}
\end{equation}
and its $k$-domain is given by $\overline{\Omega_3}.$ The right scattering coefficients $T_{\text{\rm{r}}}(k),$ $R(k),$ $N(k)$ are defined via the spacial
asymptotics of $g(k,x)$ as $x\to +\infty$ given by
\begin{equation}\label{2.15}
g(k,x)=\begin{cases}
e^{kx}\,T_{\text{\rm{r}}}(k)^{-1}[1+o(1)]+e^{zkx}R(k)\,T_{\text{\rm{r}}}(k)^{-1}[1+o(1)], \qquad k\in \mathcal L_3,\\
\noalign{\medskip}
e^{kx}\,T_{\text{\rm{r}}}(k)^{-1}[1+o(1)], \qquad k\in \Omega_3,\\
\noalign{\medskip}
e^{kx}\,T_{\text{\rm{r}}}(k)^{-1}[1+o(1)]+e^{z^2 kx}N(k)\,T_{\text{\rm{r}}}(k)^{-1}[1+o(1)],\qquad k\in \mathcal L_4,
\end{cases}
\end{equation}
where $\mathcal L_3$ and $\mathcal L_4$ are the directed half lines shown in Figure~\ref{figure2.1} corresponding to the lower and upper
boundaries of $\Omega_3.$ We remark that $\mathcal L_3$ and $\mathcal L_4$ have the respective parametric representations given by
\begin{equation*}
\mathcal L_3:=\{k\in\mathbb C: k=-zs \text{\rm{ for }}  s\in[0,+\infty)\},
\end{equation*}
\begin{equation*}
\mathcal L_4:=\{k\in\mathbb C: k=-z^2s \text{\rm{ for }}  s\in[0,+\infty)\}.
\end{equation*}
The right transmission coefficient $T_{\text{\rm{r}}}(k)$ is defined in $\overline{\Omega_3},$ the right primary reflection coefficient $R(k)$ is defined
on $\mathcal L_3,$ and the right secondary reflection coefficient $N(k)$ is defined on $\mathcal L_4.$ 

The basic solution $m(k,x)$ to \eqref{1.1} has the $k$-domain $\overline{\Omega_2},$ and it satisfies the spacial asymptotics as $x\to +\infty$ given by
\begin{equation}\label{2.18}
m(k,x)=\begin{cases}
\begin{aligned}
e^{kx}&\,T_{\text{\rm{l}}}(z^2k)^{-1}\,T_{\text{\rm{r}}}(zk)
[1+o(1)]\\
&-e^{z^2kx}M(k)\,T_{\text{\rm{l}}}(k)^{-1} \,T_{\text{\rm{r}}}(zk)
[1+o(1)], \qquad k\in \mathcal L_2,
\end{aligned}\\
\noalign{\medskip}
e^{kx}\,T_{\text{\rm{l}}}(z^2k)^{-1}\,T_{\text{\rm{r}}}(zk)[1+o(1)], \qquad k\in \Omega_2,\\
\noalign{\medskip}
e^{kx}\,T_{\text{\rm{l}}}(z^2k)^{-1}\,T_{\text{\rm{r}}}(zk)[1+o(1)],\qquad k\in \mathcal L_3,
\end{cases}
\end{equation}
and it satisfies the spacial asymptotics as $x\to -\infty$ given by
\begin{equation}\label{2.19}
m(k,x)=\begin{cases}
e^{kx}[1+o(1)], \qquad k\in \mathcal L_2,\\
\noalign{\medskip}
e^{kx}[1+o(1)],  \qquad k\in \Omega_2,\\
\noalign{\medskip}
e^{kx}[1+o(1)]-e^{zkx}R(k)\,T_{\text{\rm{r}}}(k)^{-1}\,T_{\text{\rm{r}}}(zk)[1+o(1)],\qquad k\in \mathcal L_3.
\end{cases}
\end{equation}
Similarly, the basic solution $n(k,x)$ to \eqref{1.1} has the $k$-domain $\overline{\Omega_4},$ and it satisfies the spacial asymptotics as $x\to +\infty$
given by
\begin{equation}\label{2.20}
n(k,x)=\begin{cases}
\begin{aligned}
e^{kx}&\,T_{\text{\rm{l}}}(zk)^{-1}\,T_{\text{\rm{r}}}(z^2k)[1+o(1)]\\
&-e^{zkx}L(k)\,T_{\text{\rm{l}}}(k)^{-1}\,T_{\text{\rm{r}}}(z^2k)[1+o(1)], \qquad k\in \mathcal L_1,
\end{aligned}\\
\noalign{\medskip}
e^{kx}\,T_{\text{\rm{l}}}(z k)^{-1}\,T_{\text{\rm{r}}}(z^2k)[1+o(1)],\qquad k\in \Omega_4,\\
\noalign{\medskip}
e^{kx}\,T_{\text{\rm{l}}}(z k)^{-1}\,T_{\text{\rm{r}}}(z^2k)[1+o(1)],\qquad k\in \mathcal L_4,
\end{cases}
\end{equation}
and it has the spacial asymptotics as $x\to -\infty$ given by
\begin{equation}\label{2.21}
n(k,x)=\begin{cases}
 e^{kx}[1+o(1)], \qquad k\in \mathcal L_1,\\
\noalign{\medskip}
e^{kx}[1+o(1)], \qquad k\in \Omega_4,\\
\noalign{\medskip}
e^{kx}[1+o(1)]
-e^{z^2 kx}N(k)\,T_{\text{\rm{r}}}(k)^{-1}\,T_{\text{\rm{r}}}(z^2k)[1+o(1)],\qquad k\in \mathcal L_4.
\end{cases}
\end{equation}

\begin{figure}[!ht]
     \centering
         \includegraphics[width=2.25in]{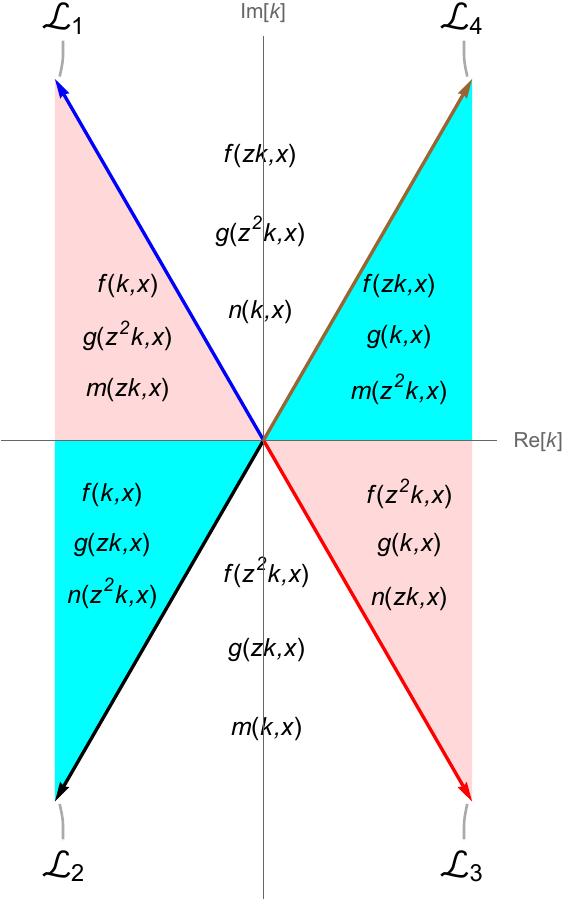}  
\caption{The $k$-domains of the basic solutions to \eqref{1.1} in each of the six regions, respectively.}
\label{figure2.2}
\end{figure}

The large $k$-asymptotics of the basic solutions $f(k,x),$ $g(k,x),$ $m(k,x),$ and $n(k,x),$ respectively, are given by
\begin{equation}\label{2.22}
f(k,x)=e^{kx}\left[1+O\left(\ds\frac{1}{k}\right)\right], \qquad  k\to\infty  \text{\rm{ in }} \overline{\Omega_1},
\end{equation}
\begin{equation}\label{2.23}
g(k,x)=e^{kx}\left[1+O\left(\ds\frac{1}{k}\right)\right], \qquad  k\to\infty  \text{\rm{ in }} \overline{\Omega_3},
\end{equation}
\begin{equation}\label{2.24}
m(k,x)=e^{kx}\left[1+O\left(\ds\frac{1}{k}\right)\right], \qquad  k\to\infty  \text{\rm{ in }} \overline{\Omega_2},
\end{equation}
\begin{equation}\label{2.25}
n(k,x)=e^{kx}\left[1+O\left(\ds\frac{1}{k}\right)\right], \qquad  k\to\infty  \text{\rm{ in }} \overline{\Omega_4}.
\end{equation}
Using the four basic solutions $f(k,x),$ $g(k,x),$ $m(k,x),$ and $n(k,x),$ we can form a fundamental set of three solutions at each $k$-value in the
complex $k$-plane, as shown in Figure~\ref{figure2.2}.

Let us briefly explain how the solutions $m(k,x)$ and $n(k,x)$ to \eqref{1.1} arise. Associated with \eqref{1.1}, we have the adjoint equation
\begin{equation}\label{2.26}
    \overline \psi'''(k,x) + \overline Q(x)\, \overline \psi'(k,x) + \overline P(x)\, \overline \psi(k,x) = k^3\,\overline \psi(k,x), \qquad x\in \mathbb R,
\end{equation}
where the adjoint potentials $\overline Q$ and $\overline P$ are defined in terms of the potentials $Q$ and $P$ as
\begin{equation}\label{2.27}
    \overline Q(x):=Q(x)^\ast, \qquad x\in \mathbb R,
\end{equation}
\begin{equation}\label{2.28}
    \overline P(x):=Q'(x)^\ast-P(x)^\ast, \qquad x\in \mathbb R.
\end{equation}
We remark that the overbar used in \eqref{2.26}--\eqref{2.28} does not denote complex conjugation, and it only refers to the quantities associated
with the adjoint equation \eqref{2.26}. We recall that an asterisk is used to denote complex conjugation and the prime in \eqref{2.28} denotes
the $x$-derivative. From \eqref{2.27} and \eqref{2.28} we observe that the adjoint potentials $\overline Q$ and $\overline P$ belong to the
Schwartz class $\mathcal S(\mathbb R)$ because $Q$ and $P$ belong to $\mathcal S(\mathbb R).$ The left and right Jost solutions to the adjoint
equation \eqref{2.26}, denoted by $\overline f(k,x)$ and $\overline g(k,x),$ respectively, are the solutions to \eqref{2.26} satisfying the respective
spacial asymptotics given in \eqref{2.10} and \eqref{2.14}, respectively. Since $k$ appears as $k^3$ in \eqref{2.26}, we see that $\overline f(zk,x),$
$\overline f(z^2k,x),$ $\overline g(zk,x),$ $\overline g(z^2k,x)$ are also solutions to \eqref{2.26}. The $k$-domains of $\overline f(k,x)$ and
$\overline g(k,x)$ coincide with the $k$-domains of $f(k,x)$ and $g(k,x),$ respectively. It is known \cite{ACTU2025,T2024} that $m(k,x)$ and $n(k,x)$ are
defined in terms of the adjoint Jost solutions, respectively, as
\begin{equation}\label{2.29}
m(k,x):=\ds\frac{T_{\text{\rm{r}}}(zk)\left[\overline f(-z^2k^\ast,x)^\ast\, ;\,\overline g(-zk^\ast,x)^\ast \right]}{z(1-z)k}, \qquad
k\in\overline{\Omega_2}, \quad x\in\mathbb R,
\end{equation}
\begin{equation}\label{2.30}
n(k,x):=\ds\frac{T_{\text{\rm{r}}}(z^2k)\left[\overline f(-zk^\ast,x)^\ast\, ;\,\overline g(-z^2 k^\ast,x)^\ast \right]}{-z(1-z)k}, 
\qquad k\in\overline{\Omega_2}, \quad x\in\mathbb R,
\end{equation}
where we recall that $T_{\text{\rm{r}}}(k)$ is the right transmission coefficient appearing in \eqref{2.15}.
We remark that in \eqref{2.29} and \eqref{2.30} we use the $2$-Wronskian of the adjoint Jost solutions, where we have defined
\begin{equation*}
\left[F(x);G(x)\right]:=F(x)\,G'(x)-F'(x)\,G(x).
\end{equation*}
In comparing $m(k,x)$ and $\left[\overline f(-z^2k^\ast,x)^\ast\, ;\,\overline g(-zk^\ast,x)^\ast \right],$ we remark that the former solution
to \eqref{1.1} has simpler spacial asymptotics and the latter quantity is analytic
in $k\in\Omega_2$ and continuous $k\in\overline{\Omega_2}.$ Similarly, in comparing the solutions $n(k,x)$ and
$\left[\overline f(-zk^\ast,x)^\ast\, ;\,\overline g(-z^2 k^\ast,x)^\ast \right]$ to \eqref{1.1}, we remark that the former has simpler spacial asymptotics
and the latter solution has the advantage that it is analytic in $k\in\Omega_4$ and continuous $k\in\overline{\Omega_4}.$ 

Let us define the 3-Wronskian of three functions $F(k,x),$ $G(k,x),$ and $H(k,x)$ as
\begin{equation*}
\left[F(x);G(x);H(x)\right]:=\begin{vmatrix}
F(x) & G(x) & H(x)\\
F'(x) & G'(x) & H'(x)\\
F''(x) & G''(x) & H''(x)
\end{vmatrix},
\end{equation*}
where the right-hand side consists of the determinant of the relevant $3\times 3$ matrix. We recall that the $3$-Wronskian of any three solutions to 
\eqref{1.1} at a $k$-value is zero if and only if those three solutions are linearly dependent. Furthermore, since the term $\psi''$ is absent in \eqref{1.1}, 
the $3$-Wronskian of any three solutions to \eqref{1.1} is independent of $x$ and hence its value can be evaluated at any
$x$-value. With the help of the spacial asymptotics in \eqref{2.10}--\eqref{2.21}, we obtain the $3$-Wronskians
\begin{equation}\label{2.33}
\left[f(k,x);g(zk,x);n(z^2k,x)\right]=-3 z(1-z)k^3 \,T_{\text{\rm{l}}}(k)^{-1},
\qquad k\in\overline{\Omega_1^\text{\rm{down}}},
\end{equation}
\begin{equation}\label{2.34}
\left[f(k,x);g(z^2k,x);m(zk,x)\right]=3(1-z^2)k^3 \,T_{\text{\rm{l}}}(k)^{-1},
\qquad k\in\overline{\Omega_1^\text{\rm{up}}}.
\end{equation}
Thus, from \eqref{2.33} we observe that the solutions $f(k,x),$ $g(zk,x),$ $n(z^2k,x)$ to \eqref{1.1} are linearly independent in $k\in\overline{\Omega_1^\text{\rm{down}}}\setminus\{0\}$
except at the $k$-values corresponding to the poles of $T_{\text{\rm{l}}}(k)$ there. Similarly, from \eqref{2.34} we see that the
solutions $f(k,x),$ $g(z^2k,x),$ $m(zk,x)$ to \eqref{1.1}
are linearly independent in $k\in\overline{\Omega_1^\text{\rm{up}}}\setminus\{0\}$ except at the $k$-values corresponding to the poles of $T_{\text{\rm{l}}}(k)$ there.
We refer the reader to \cite{ACTU2025,T2024} for further properties of the basic solutions $f(k,x),$ $g(k,x),$ $m(k,x),$ $n(k,x)$ to \eqref{1.1} and
the basic properties of
the six scattering coefficients $T_{\text{\rm{l}}}(k),$ $L(k),$ $M(k),$ $T_{\text{\rm{r}}}(k),$ $R(k),$ $N(k).$

The scattering coefficients for \eqref{1.1} satisfy various properties. For example, the left reflection coefficients $L(k)$ and $M(k)$ are related to the
transmission coefficients $T_{\text{\rm{l}}}(k)$ and $T_{\text{\rm{r}}}(k)$ as
     \begin{equation}\label{2.35}
         T_{\text{\rm{r}}}(z^2k)^{-1}= T_{\text{\rm{l}}}(k)^{-1}\, T_{\text{\rm{l}}}(zk)^{-1}\, [1-L(k)\, M(zk)], \qquad k\in\mathcal L_1.
     \end{equation}
Similarly, the right reflection coefficients $R(k)$ and $N(k)$ are related to the transmission coefficients $T_{\text{\rm{l}}}(k)$ and $T_{\text{\rm{r}}}(k)$ 
as 
 \begin{equation}\label{2.36}
          T_{\text{\rm{l}}}(z^2k)^{-1} = T_{\text{\rm{r}}}(k)^{-1}\, T_{\text{\rm{r}}}(zk)^{-1}\,\left[1 - R(k)\, N(zk)\right], \qquad k\in\mathcal L_3.
      \end{equation} 
For the proof of \eqref{2.35} and \eqref{2.36}, we refer the reader to Theorem~3.3.3 of \cite{T2024}. We remark that when $M(k)\equiv 0,$ from 
\eqref{2.35} we get
   \begin{equation}\label{2.37}
         T_{\text{\rm{r}}}(z^2k)^{-1}= T_{\text{\rm{l}}}(k)^{-1}\, T_{\text{\rm{l}}}(zk)^{-1}, \qquad k\in\mathcal L_1.
     \end{equation}
Similarly, when $N(k)\equiv 0,$ from \eqref{2.36} we obtain
 \begin{equation}\label{2.38}
          T_{\text{\rm{l}}}(z^2k)^{-1} = T_{\text{\rm{r}}}(k)^{-1}\, T_{\text{\rm{r}}}(zk)^{-1}, \qquad k\in\mathcal L_3.
      \end{equation}

\section{The bound states}
\label{section3}

A bound-state solution to \eqref{1.1} corresponds to a nontrivial solution which is square integrable in $x\in\mathbb R.$
Let us assume that \eqref{1.1} has a bound state at the $k$-value $k_j$ in the complex $k$-plane. Then, \eqref{1.1} at $k=k_j,$ i.e. the equation
\begin{equation}
\label{3.1}
\psi'''+Q(x)\,\psi'+P(x)\,\psi=k_j^3\,\psi, \qquad x\in\mathbb R,
\end{equation}
has a solution $\psi(k_j,x)$ satisfying
\begin{equation*}
\int_{-\infty}^\infty \,dx \left| \psi(k_j,x)\right|^2<+\infty.
\end{equation*}
The solution $\psi(k_j,x)$ is a bound-state solution to \eqref{1.1} at $k=k_j.$ Since \eqref{3.1} is a linear homogeneous equation, any constant
multiple of a bound-state solution is also a bound-state solution. The number of linearly independent bound-state solutions to \eqref{3.1}
determines the multiplicity of the bound state at $k=k_j.$ In this paper, we only consider simple bound states, i.e. bound states of multiplicity
one. We refer the reader to \cite{AE2019,AE2022,AEU2023} for the analysis of nonsimple bound states, i.e. bound states with multiplicity greater
than one.
We refer the reader to the analysis in Section~3 of \cite{ACTU2025} for a detailed description of the bound states for \eqref{1.1}. In the present
paper, we provide a brief summary of the bound-state analysis relevant to the formulation of the inverse scattering problem analyzed in
Sections~\ref{section4} and \ref{section5}.

Suppose that the bound state at $k=k_j$ with $k_j\neq 0$ occurs when the argument of $k_j$ lies in the interval $[7\pi/6,4\pi/3).$ In that case
we have $T_{\text{\rm{l}}}(k_j)^{-1}=0,$ where we recall that $T_{\text{\rm{l}}}(k)$ is the left reflection coefficient appearing in \eqref{2.11}.
We then observe that the right-hand side of \eqref{2.33} becomes zero, indicating that the three solutions $f(k_j,x),$ $g(zk_j,x),$ and $n(z^2k_j,x)$ to
\eqref{3.1} are linearly dependent. The analysis in Section~3 of \cite{ACTU2025} shows that, when
$T_{\text{\rm{l}}}(k_j)^{-1}=0$
with 
$\arg[k_j]\in[7\pi/6,4\pi/3),$ the solution $n(z^2k_j,x)$ to \eqref{3.1} does not vanish as $x\to+\infty$ but $f(k_j,x)$ and $g(zk_j,x)$ both vanish exponentially as
$x\to\pm\infty.$ Thus, there exists a nonzero constant $D(k_j)$ such that
\begin{equation}\label{3.3}
f(k_j,x)=D(k_j) \, g(zk_j,x),\qquad \arg[k_j]\in \left[\ds\frac{7\pi}{6},\ds\frac{4\pi}{3}\right), \quad x\in\mathbb R.
\end{equation}
We refer to the nonzero complex constant $D(k_j)$ as the dependency constant at the bound state at $k=k_j.$ The constant $D(k_j)$ is the 
analog of the bound-state dependency constant defined in (1.3) of \cite{AK2001} for the full-line Schr\"odinger equation. As in (1.3) of \cite{AK2001}, 
the bound-state dependency constant corresponds to the ratio of the left Jost solution to the right Jost solution at the bound state.

From \eqref{3.3} we observe that each of $f(k_j,x)$ and $g(zk_j,x)$ are square-integrable solutions to \eqref{3.1}. By introducing the positive
constant $c_{\text{\rm{l}}}(k_j)$ via
\begin{equation*}
c_{\text{\rm{l}}}(k_j):=\ds\frac{1}{\sqrt{\ds\int_{-\infty}^\infty dx \left| f(k_j,x) \right|^2}}, \qquad \arg[k_j]\in\left[\ds\frac{7\pi}{6},\ds\frac{4\pi}{3}\right),
\end{equation*}     
we observe that the quantity $c_{\text{\rm{l}}}(k_j) f(k_j,x)$ is a normalized square-integrable solution to \eqref{3.1}, i.e. we have 
\begin{equation*}
\int_{-\infty}^\infty dx \left| c_{\text{\rm{l}}}(k_j)\,f(k_j,x)\right|^2=1.
\end{equation*}   
Since $f(k_j,x)$ is the left Jost solution to
\eqref{3.1}, it is appropriate to refer to $c_{\text{\rm{l}}}(k_j)$ as the left bound-state normalization constant, which is the analog of the left
bound-state normalization constant appearing in (1.7) of \cite{AK2001} for the full-line Schr\"odinger equation. Similarly, we can introduce the right
bound-state normalization constant $c_{\text{\rm{r}}}(k_j)$ via 
\begin{equation*}
c_{\text{\rm{r}}}(k_j):=
\ds\frac{1}{\sqrt{\ds\int_{-\infty}^\infty dx\, \left| g(zk_j,x)\right|^2}}, \qquad \arg[k_j]\in\left[\ds\frac{7\pi}{6},\ds\frac{4\pi}{3}\right).
\end{equation*}
Then, the quantity $c_{\text{\rm{r}}}(k_j) g(zk_j,x)$ corresponds to a normalized bound-state solution to \eqref{1.1} at $k=k_j.$ 

In case we have a bound state occurring at a nonzero $k$-value with $\arg[k]\in(2\pi/3,5\pi/6],$ then we can assume that this particular
$k$-value occurs at $k=k_j^\ast$ with $\arg[k_j]\in(7\pi/6,4\pi/3],$ where we recall that an asterisk denotes complex conjugation.
Since the points $k_j$ and $k_j^\ast$ are located symmetrically with respect to the negative real axis in the complex $k$-plane, the bound-state
analysis at $k=k_j$ can be used to describe the bound-state analysis at $k=k_j^\ast$ with $\arg[k_j^\ast]\in(2\pi/3,5\pi/6].$ Thus, at the bound 
state occurring at $k=k_j^\ast,$ we have  $T_{\text{\rm{l}}}(k_j^\ast)^{-1}=0.$ Furthermore, the solutions $f(k_j^\ast,x)$ and $g(z^2k_j^\ast,x)$
to \eqref{1.1} at $k=k_j^\ast$ become linearly dependent, and they are related to each other as   
\begin{equation*}
f(k_j^\ast,x)=
D(k_j^\ast)\, g(z^2k_j^\ast,x), \qquad \arg[k_j^\ast]\in\left(\ds\frac{2\pi}{3},\ds\frac{5\pi}{6}\right],
\end{equation*}
where $D(k_j^\ast)$ is a nonzero complex constant, which we call the dependency constant at the bound state at $k=k_j^\ast.$

For the analysis of the bound states occurring at $k$-values in the sector $\arg[k]\in(5\pi/6,\pi)$ or in the sector $\arg[k]\in(\pi,7\pi/6),$ we refer the reader
to Section~3 of \cite{ACTU2025}, where the dependency constants at those $k$-values are described.

\section{The case of zero secondary reflection coefficients}
\label{section4}

In this section we consider the important special case for \eqref{1.1} when the secondary reflection coefficients $M(k)$ and $N(k)$ are both zero.
This special case is relevant in introducing a properly posed Riemann--Hilbert problem in the complex $k$-plane and obtaining the associated
linear integral equation, which is the analog of the Marchenko integral equation \cite{CS1989,DT1979,F1967,L1987,M2011} available arising in
the analysis of the inverse scattering 
theory for the full-line Schr\"odinger equation. 

In the special case when the left secondary reflection coefficient $M(k)$ is zero, we have \cite{ACTU2025,T2024}
\begin{equation}\label{4.1}
T_{\text{\rm{l}}}(k)\,f(k,x)=m(k,x),\qquad x\in\mathbb R,\quad k\in\mathcal L_2.
\end{equation}
The following argument shows why we have \eqref{4.1} when $M(k)\equiv 0.$ In order to show that \eqref{4.1} holds when $M(k)\equiv 0,$ it is enough to
show that both $m(k,x)$ and $T_{\text{\rm{l}}}(k)\,f(k,x)$ satisfy \eqref{1.1} for $k\in\mathcal L_2$ and they both have the same spacial asymptotics
as $x\to -\infty$ for $k\in\mathcal L_2.$ Since \eqref{1.1} is a linear homogeneous equation and both $m(k,x)$ and $f(k,x)$ are solutions to
\eqref{1.1}, it is clear that $m(k,x)$ and $T_{\text{\rm{l}}}(k) f(k,x)$ are both solutions to \eqref{1.1}. Furthermore, comparing the third line of
\eqref{2.11} with $M(k)\equiv 0$ and the first line of \eqref{2.19}, we see that each of $m(k,x)$ and $T_{\text{\rm{l}}}(k)\,f(k,x)$ behaves as
$e^{kx}[1+o(1)]$ as $x\to -\infty$ for $k\in\mathcal L_2.$ Thus, \eqref{4.1} holds when $M(k)\equiv 0.$ 

Similarly, when the right secondary reflection coefficient $N(k)$ is zero, we have \cite{ACTU2025,T2024}
\begin{equation}\label{4.2}
g(k,x)=n(k,x),\qquad x\in\mathbb R,\quad k\in\mathcal L_4.
\end{equation}
The following argument shows why \eqref{4.2} holds when $N(k)\equiv 0.$ For \eqref{4.2} to hold, it is sufficient to show that each side of \eqref{4.2}
is a solution to \eqref{1.1} and has the same spacial asymptotics as $x\to +\infty$ for $k\in\mathcal L_4.$ We already know that both $f(k,x)$ and
$n(k,x)$ are solution to \eqref{1.1} when $k\in\mathcal L_4.$ When $N(k)\equiv 0,$ from the first line of \eqref{2.14} and from the third line of
\eqref{2.21}, we see that each of $g(k,x)$ and $n(k,x)$ behaves as $e^{kx}[1+o(1)]$ as $x\to -\infty$ for $k\in\mathcal L_4.$ Thus, \eqref{4.2} holds when
$N(k)\equiv 0.$ 
 
When both secondary reflection coefficients are zero, with the help of \eqref{4.1} and \eqref{4.2} we introduce the solution $\Phi_+(k,x)$ to \eqref{1.1}
with the $k$-domain $\overline{\Omega_1}\cup\overline{\Omega_2}$ and the solution $\Phi_-(k,x)$ to \eqref{1.1} with the $k$-domain
$\overline{\Omega_3}\cup\overline{\Omega_4},$ where we have defined
\begin{equation}\label{4.3}
    \Phi_+(k,x) := 
    \begin{cases}
        T_\text{\rm{l}}(k)\, f(k,x), \qquad k\in \overline{\Omega_1},\\
                \noalign{\medskip}
        m(k,x), \qquad k\in \overline{\Omega_2},
    \end{cases}
\end{equation}
\begin{equation}\label{4.4}
    \Phi_-(k,x) := 
    \begin{cases}
        g(k,x), \qquad k\in \overline{\Omega_3},\\
        \noalign{\medskip}
        n(k,x), \qquad k\in \overline{\Omega_4}.
    \end{cases}
\end{equation}
From \eqref{4.1} and \eqref{4.3}, we conclude that the solution $m(k,x)$ to \eqref{1.1} has a meromorphic continuation from $\Omega_2$
to $\Omega_1,$ or equivalently the meromorphic function $T_{\text{\rm{l}}}(k)\,f(k,x)$ defined in $k\in\Omega_1$ has a meromorphic continuation to
$\overline{\Omega_2}.$ Similarly, from \eqref{4.2} and \eqref{4.4}, we see that $g(k,x)$ has a meromorphic extention from $\Omega_3$
to $\Omega_4.$ or equivalently $n(k,x)$ has an analytic continuation from $\Omega_4$ to $\Omega_3.$

\begin{figure}[!ht]
     \centering
         \includegraphics[width=2.in,height=3.4in]{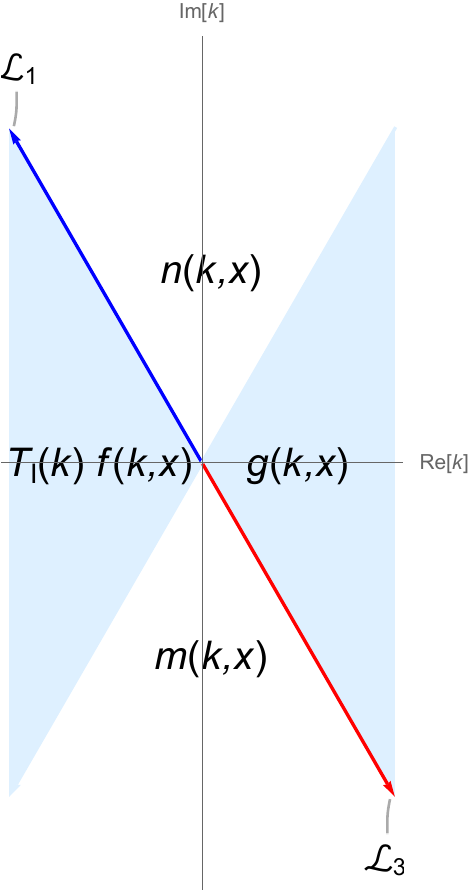}      \hskip .5in
         \includegraphics[width=2.in,height=3.4in]{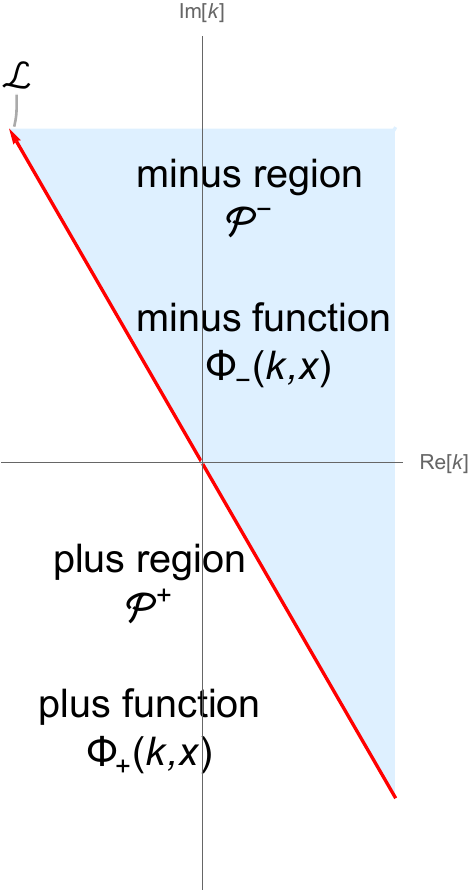} 
\caption{The $k$-domains of $T_{\text{\rm{l}}}(k) f(k,x),$ $m(k,x),$ $g(k,x),$ and $n(k,x),$ respectively, are shown on the left
plot. The right plot shows the plus and minus regions in the complex $k$-plane separated by the directed full line $\mathcal L,$
as well as the plus and minus functions in their respective $k$-domains.}
\label{figure4.1}
\end{figure}

Let us define the full directed line $\mathcal L$ as the union of the half lines $\mathcal L_1$ and $-\mathcal L_3,$ where $-\mathcal L_3$
denotes $\mathcal L_3$ but with its direction reversed. Thus, $\mathcal L$ has the parametrization given by
\begin{equation*}
\mathcal L:=\{k\in\mathbb C: k=zs \text{\rm{ for }}  s\in(-\infty,+\infty)\}.
\end{equation*}
The line $\mathcal L$ separates the complex $k$-plane into two open half planes $\mathcal P^+$ and $\mathcal P^-,$ respectively, as shown on the right
plot of Figure~\ref{figure4.1}. We note that the open half plane $\mathcal P^+$ corresponds to the interior of
$\overline{\Omega_1}\cup\overline{\Omega_2}$ and the open half plane $\mathcal P^-$ corresponds to the interior of
$\overline{\Omega_3}\cup\overline{\Omega_4}.$ From \eqref{4.3}, we observe that, for each fixed $x\in\mathbb R,$ the quantity $\Phi_+(k,x)$ is
meromorphic in $k\in\mathcal P^+.$ Similarly, from \eqref{4.4}, we observe that, for each fixed $x\in\mathbb R,$ the quantity $\Phi_-(k,x)$ is
meromorphic in $k\in\mathcal P^-.$ We refer to $\mathcal P^+$ as the plus region and $\mathcal P^-$ as the minus region. We use
$\overline{\mathcal P^+}$ to denote the closure of $\mathcal P^+,$ i.e. we let $\overline{\mathcal P^+}:=\mathcal P^+\cup\mathcal L.$ Similarly,
we use $\overline{\mathcal P^-}$ to denote the closure of $\mathcal P^-,$ i.e. we let $\overline{\mathcal P^-}:=\mathcal P^-\cup\mathcal L.$

Using $\Phi_+(k,x)$ and $\Phi_-(k,x)$ we are interested in the formulation of a Riemann--Hilbert problem for \eqref{1.1} when the secondary
reflection coefficients $M(k)$ and $N(k)$ are both zero. This Riemann--Hilbert problem is posed on the complex $k$-plane separated by the full
directed line $\mathcal L$ as shown on the right plot of Figure~\ref{figure4.1}. We refer to $\Phi_+(k,x)$ as the plus function with the $k$-domain
$\overline{\mathcal P^+}.$ Similarly, we refer to $\Phi_-(k,x)$ as the minus function with the $k$-domain $\overline{\mathcal P^-}.$ 
In the formulation of our Riemann--Hilbert problem, the complex $k$-plane is divided into two regions by one single full line. We refer
to such a formulation as a proper formulation of the Riemann--Hilbert problem. In case the boundary separating the plus and minus regions were not a single
full line, then we would refer to the formulation of that Riemann--Hilbert problem as an improper formulation.

Our Riemann--Hilbert problem relevant to \eqref{1.1} arises as follows. The complex $k$-plane is separated by the directed full line $\mathcal L$ into
the two regions $\mathcal P^+$ and $\mathcal P^-.$ According to the commonly used convention, the region to the left of the directed line
$\mathcal L$ is the plus region $\mathcal P^+$ and the region to the right of $\mathcal L$ is the minus region $\mathcal P^-.$
In a standard Riemann--Hilbert problem, we would have a sectionally analytic function in $k\in\mathbb C$ with a jump on $\mathcal L.$
In other words, for each fixed $x\in\mathbb R,$ we would have the sectionally analytic function $\Phi(k,x),$ where $\Phi_+(k,x)$ and $\Phi_-(k,x)$
make up the two sections of $\Phi(k,x)$ in such a way that $\Phi_+(k,x)$ is analytic in $k\in\mathcal P^+$ and $\Phi_-(k,x)$ is analytic in
$k\in\mathcal P^-.$ In our own formulation of the Riemann--Hilbert problem, we deal with a sectionally meromorphic function $\Phi(k,x),$ where the 
section $\Phi_+(k,x)$ is meromorphic in $k\in\mathcal P^+$ and the section $\Phi_-(k,x)$ is meromorphic in $k\in\mathcal P^-.$ As in the standard 
Riemann--Hilbert problem we choose $\Phi_+(k,x)$ and $\Phi_-(k,x)$ so that $\Phi_+(k,x)$ vanishes as $k\to\infty$ in $\overline{\mathcal P^+}$ 
and $\Phi_-(k,x)$ vanishes as $k\to\infty$ in $\overline{\mathcal P^-}.$ That normalization is chosen so that
the Riemann-Hilbert problem has a unique solution when an
appropriate input data set is provided to solve the Riemann--Hilbert problem. The solution to the Riemann--Hilbert problem is obtained by constructing
$\Phi_+(k,x)$ and $\Phi_-(k,x)$ when their difference $\Phi_+(k,x)-\Phi_-(k,x)$ is known for $k\in\mathcal L.$ In other words, knowing the jump 
on $\mathcal L$ for the sectionally meromorphic function $\Phi(k,x),$ we would like to construct $\Phi(k,x)$ when $k\in\mathbb C.$

When the secondary reflection coefficients $M(k)$ and $N(k)$ are both zero, the functions $\Phi_+(k,x)$ and $\Phi_-(k,x),$ defined in \eqref{4.3} 
and \eqref{4.4}, respectively satisfy \cite{T2024} the jump condition on $\mathcal L$ given by
\begin{equation}\label{4.6}
\Phi_+(k,x)-\Phi_-(k,x)=J(k,x), \qquad k\in\mathcal L.
\end{equation}
The jump $J(k,x)$ appearing on the right-hand side of \eqref{4.6} is expressed in terms of the Jost solutions $f(k,x)$ and $g(k,x)$ and the
remaining scattering coefficients $T_{\text{\rm{l}}}(k),$ $T_{\text{\rm{r}}}(k),$ $L(k),$ and $R(k)$ as 
\begin{equation}\label{4.7}
J(k,x)=\begin{cases}
L(k)\,T_{\text{\rm{l}}}(zk)\,f(zk,x), \qquad k\in\mathcal L_1,\\
\noalign{\medskip}
-\ds\frac{R(k)\,T_{\text{\rm{r}}}(zk)\,g(zk,x)}{T_{\text{\rm{r}}}(k)}, \qquad k\in -\mathcal L_3.
\end{cases}
\end{equation}
We recall that $-\mathcal L_3$ appearing in \eqref{4.7} corresponds to the half line $\mathcal L_3$ with the direction reversed.

In the following we briefly describe how \eqref{4.7} is obtained when $M(k)\equiv 0$ and $N(k) \equiv 0.$ For the details of the derivation,
we refer the reader to the proof of Theorem~4.0.4 of \cite{T2024}. In order to establish the first line of \eqref{4.7}, as seen from \eqref{4.3} and
\eqref{4.4}, we need to show that
\begin{equation}\label{4.8}
T_{\text{\rm{l}}}(k)\,f(k,x)-n(k,x)=L(k)\,T_{\text{\rm{l}}}(zk)\,f(zk,x), \qquad k\in\mathcal L_1.
\end{equation}
For this, it is sufficient to prove that each side of \eqref{4.8} is a solution to \eqref{1.1} satisfying the same spacial asymptotics as $x\to +\infty$
for $k\in\mathcal L_1.$ We note that \eqref{1.1} is a linear homogeneous differential equation, and furthermore $f(k,x),$ $n(k,x),$ and $f(zk,x)$ are
solutions to \eqref{1.1}. Thus, it follows that each side of \eqref{4.8} is a solution to \eqref{1.1}. From the first line of \eqref{2.10} we see that the
right-hand side of \eqref{4.8} has the spacial asymptotics as $x\to +\infty$ given by
\begin{equation}\label{4.9}
L(k)\,T_{\text{\rm{l}}}(zk)\,f(zk,x)=L(k)\,T_{\text{\rm{l}}}(zk)\,e^{zkx}\left[1+o(1)\right], \qquad k\in\mathcal L_1.
\end{equation}
On the other hand, from the first line of \eqref{2.10} and the first line of \eqref{2.20} we see that the left-hand side of \eqref{4.8} has the spacial
asymptotics as $x\to +\infty$ given by
\begin{equation}\label{4.10}
\begin{aligned}
T_{\text{\rm{l}}}(k)f(k,x)-n(k,x)=&e^{kx}\left[T_{\text{\rm{l}}}(k)^{-1}-T_{\text{\rm{l}}}(zk)^{-1}T_{\text{\rm{r}}}(z^2k)\right][1+o(1)]\\
                                                   &+e^{zkx}L(k)\,T_{\text{\rm{l}}}(k)^{-1}T_{\text{\rm{r}}}(z^2k)[1+o(1)], \qquad k\in\mathcal L_1.
\end{aligned}                                                  
\end{equation}
From \eqref{2.37}, we know that when $M(k)\equiv 0,$ the coefficient of $e^{kx}$ on the right-hand side of \eqref{4.10} vanishes and \eqref{4.10} reduces to the spacial
asymptotics as $x\to +\infty$ given by
\begin{equation}\label{4.11}
T_{\text{\rm{l}}}(k)\,f(k,x)-n(k,x)=e^{zkx}L(k)\,T_{\text{\rm{l}}}(k)^{-1}T_{\text{\rm{r}}}(z^2k)[1+o(1)], \qquad k\in\mathcal L_1.
\end{equation}
We observe that the right-hand sides of \eqref{4.9} and \eqref{4.11} coincide as a consequence of \eqref{2.37}. Thus, we have proved that
\eqref{4.8} holds. To complete the proof of \eqref{4.7}, as seen  from \eqref{4.3}--\eqref{4.7}, we also need to show that
\begin{equation}\label{4.12}
m(k,x)-g(k,x)=-\ds\frac{R(k)\,T_{\text{\rm{r}}}(zk)\,g(zk,x)}{T_{\text{\rm{r}}}(k)}, \qquad k\in\mathcal L_3.
\end{equation}
For the proof of \eqref{4.12}, we proceed as follows. We note that each side of \eqref{4.12} is a solution to \eqref{1.1} due to the fact that \eqref{1.1} is
a linear homogeneous differential equation and that $m(k,x),$ $g(k,x),$ and $g(zk,x)$ are all solutions to \eqref{1.1}. Thus, to establish \eqref{4.12}
it is sufficient to prove that each side of \eqref{4.12} has the same spacial asymptotics as $x\to -\infty.$ From the first line of \eqref{2.14}, we see that
the right-hand side of \eqref{4.12} has the spacial asymptotics as $x\to -\infty$ given by
\begin{equation}\label{4.13}
-\ds\frac{R(k)\,T_{\text{\rm{r}}}(zk)\,g(zk,x)}{T_{\text{\rm{r}}}(k)}=-\ds\frac{R(k)\,T_{\text{\rm{r}}}(zk)\,e^{zkx}[1+o(1)]}{T_{\text{\rm{r}}}(k)}, \qquad k\in\mathcal L_3.
\end{equation}
On the other hand, from the first line to \eqref{2.14} and the third line of \eqref{2.19} we observe that the left-hand side of \eqref{4.12} has the spacial
asymptotics as $x\to -\infty$ given by
\begin{equation}\label{4.14}
m(k,x)-g(k,x)=-e^{zkx}R(k)\,T_{\text{\rm{r}}}(k)^{-1}\,T_{\text{\rm{r}}}(zk), \qquad k\in\mathcal L_3.
\end{equation}
Hence, the right-hand sides of \eqref{4.13} and \eqref{4.14} agree and consequently \eqref{4.12} holds. 

With the help of \eqref{4.6} and \eqref{4.7}, we formulate our
Riemann--Hilbert problem as follow, by recalling that we assume that the secondary reflection coefficients $M(k)$ and $N(k)$ for \eqref{1.1} are
both zero. We are given the left scattering coefficients $T_{\text{\rm{l}}}(k)$ and $L(k)$ for $k\in\mathcal L_1.$ Similarly, we are given the right
scattering coefficients $T_{\text{\rm{r}}}(k)$ and $R(k)$ for $k\in\mathcal L_3.$ Furthermore, we know the poles of the extension of
$T_{\text{\rm{l}}}(k)$ from $k\in\mathcal L_1$ to $k\in\Omega_1.$ For each fixed $x\in\mathbb R,$ we would like to determine the function
$\Phi_+(k,x),$ which is meromorphic in $k\in\mathcal P^+$ with the poles coinciding with the poles of $T_{\text{\rm{l}}}(k)$ in $\Omega_1$ in such
a way that $\Phi_+(k,x)$ vanishes as $k\to\infty$ in $\mathcal P^+.$ For each fixed $x\in\mathcal R,$ we also would like to determine the function
$\Phi_-(k,x),$ which is meromorphic in $k\in\mathcal P^-$ and vanishing as $k\to\infty$ in $\overline{\mathcal P^-}.$

The solution to the aforementioned Riemann--Hilbert problem yields the solution to the inverse scattering problem for \eqref{1.1}. This can be seen
as follows. Once we recover $\Phi_+(k,x)$ from the input scattering data set, as seen from \eqref{4.3}, we have the solutions $f(k,x)$ and
$n(k,x)$ to \eqref{1.1} with their respective $k$-domains $\overline{\Omega_1}$ and $\overline{\Omega_2}.$ For example, using $f(k,x)$ in
\eqref{1.1} we can recover the potentials $Q$ and $P$ appearing in \eqref{1.1}. The recovery of these two potentials can be accomplished by using $f(k,x),$
$f'(k,x),$ and $f(k,x)$ in
\eqref{1.1}. Alternatively, if we have the solution $\Phi_-(k,x)$ to our Riemann--Hilbert problem, as seen from \eqref{4.4}, we get the solutions
$g(k,x)$ and $n(k,x)$ to \eqref{1.1} in their respective $k$-domains $\overline{\Omega_3}$ and $\overline{\Omega_4}.$ We can then recover the
potentials $Q$ and $P$ by using either of $g(k,x)$ and $n(k,x)$ and their
$x$-derivatives in \eqref{1.1}.

The recovery of the potentials $Q$ and $P$ from the left Jost solution $f(k,x)$ can also be achieved by exploiting the large $k$-asymptotics of
$f(k,x)$ as follows. It is known \cite{ACTU2025, T2024} that we have
\begin{equation}\label{4.15}
 f(k,x) = e^{kx}\left[1 +\ds\frac{u_1(x)}{k} + \ds\frac{u_2(x)}{k^2} + O\left(\ds\frac{1}{k^3}\right)\right],\qquad k\to\infty  \text{\rm{ in }} \overline{\Omega_1},\quad x\in\mathbb R, 
 \end{equation}
where we have let
\begin{equation}\label{4.16}
 u_1(x):= \ds\frac{1}{3}\int_x^{\infty} dy\, Q(y),\qquad x\in\mathbb R,
 \end{equation}
 \begin{equation}\label{4.17}
 u_2(x):= -\ds\frac{1}{3}\int_x^{\infty} dy \left[Q'(y)-P(y)\right] + \ds\frac{1}{18}\left[\int_x^{\infty} dy\, Q(y)\right]^2,
 \qquad x\in\mathbb R.
  \end{equation}
From \eqref{4.16} and \eqref{4.17}, we recover $Q$ and $P$ as
\begin{equation}\label{4.18}
 Q(x)= -3\,\ds\frac{du_1(x)}{dx}, \qquad x\in\mathbb R,
 \end{equation}
 \begin{equation}\label{4.19}
 P(x)= 3\left[u_1(x)\,\ds\frac{du_1(x)}{dx}- \ds\frac{d^2u_1(x)}{dx^2}-\ds\frac{du_2(x)}{dx}\right], \qquad x\in\mathbb R.
 \end{equation}

Alternatively, the potentials $Q$ and $P$ can be recovered from the large $k$-asymptotics of the right Jost solution $g(k,x)$ as follows.
We have \cite{ACTU2025,T2024} the asymptotics 
\begin{equation}\label{4.20}
 g(k,x) = e^{kx}\left[1 +\ds\frac{v_1(x)}{k} + \ds\frac{v_2(x)}{k^2} + O\left(\ds\frac{1}{k^3}\right)\right],\qquad k\to\infty  \text{\rm{ in }}\overline{\Omega_3}, 
 \end{equation}
where we have let
\begin{equation}\label{4.21}
 v_1(x):= -\ds\frac{1}{3}\int_{-\infty}^x dy\, Q(y),\qquad x\in\mathbb R,
 \end{equation}
 \begin{equation}\label{4.22}
 v_2(x):= \ds\frac{1}{3}\int_{-\infty}^x dy \left[Q'(y)-P(y)\right] + \ds\frac{1}{18}\left[\int_{-\infty}^x dy\, Q(y)\right]^2, \qquad x\in\mathbb R.
  \end{equation}
Using \eqref{4.21} and \eqref{4.22}, we recover $Q$ and $P$ as
\begin{equation}\label{4.23}
 Q(x)= -3\,\ds\frac{dv_1(x)}{dx}, \qquad x\in\mathbb R,
 \end{equation}
 \begin{equation}\label{4.24}
 P(x)= 3\left[v_1(x)\,\ds\frac{dv_1(x)}{dx}- \ds\frac{d^2v_1(x)}{dx^2}-\ds\frac{dv_2(x)}{dx}\right], \qquad x\in\mathbb R.
 \end{equation}

Having described how we recover the potentials $Q$ and $P$ from the Jost solutions $f(k,x)$ and $g(k,x),$ let us explain how to obtain those
potentials from the solutions $\Phi_+(k,x)$ and $\Phi_-(k,x)$ to the Riemann--Hilbert problem posed in \eqref{4.6}. Since we already know
$T_{\text{\rm{l}}}(k)$ in $\overline{\Omega_1}$ as a part of the input data set for our Riemann--Hilbert problem,
we also know its large $k$-asymptotics given by
\begin{equation}\label{4.25}
T_{\text{\rm{l}}}(k)=1+\ds\frac{t_{\text{\rm{l}}1}}{k}+\ds\frac{t_{\text{\rm{l}}2}}{k^2}+O\left(\ds\frac{1}{k^3}\right),\qquad k\to\infty  \text{\rm{ in }}\overline{\Omega_1}, 
 \end{equation}
where $t_{\text{\rm{l}}1}$ and $t_{\text{\rm{l}}2}$ are some complex constants. From the first line of \eqref{4.3}, with the help of \eqref{4.15} and
\eqref{4.25} we obtain
 \begin{equation}\label{4.26}
e^{-kx}\,\Phi_+(k,x)=1+\ds\frac{t_{\text{\rm{l}}1}+u_1(x)}{k}+\ds\frac{t_{\text{\rm{l}}2}+t_{\text{\rm{l}}1}\,u_1(x)+u_2(x)}{k^2}+O\left(\ds\frac{1}{k^3}\right),\qquad k\to\infty  \text{\rm{ in }}\overline{\Omega_1}.
 \end{equation}
Thus, we can obtain $u_1(x)$ and $u_2(x)$ from the asymptotics of $\Phi_+(k,x)$ as $k\to +\infty$ in $\overline{\Omega_1},$ and by using
those quantities in \eqref{4.18} and \eqref{4.19} we recover $Q$ and $P.$

In a similar manner, we can recover $Q$ and $P$ directly from the solution $\Phi_-(k,x)$ to the Riemann--Hilbert problem by proceeding as follows.
From the first line of \eqref{4.4} we see that $\Phi_-(k,x)$ and $g(k,x)$ coincide when $k\in\overline{\Omega_3}.$ 
Hence, with the help of \eqref{4.20} we have
 \begin{equation}\label{4.27}
e^{-kx}\,\Phi_-(k,x)=1+\ds\frac{v_1(x)}{k}+\ds\frac{v_2(x)}{k^2}+O\left(\ds\frac{1}{k^3}\right),\qquad k\to\infty  \text{\rm{ in }}\overline{\Omega_3}.
 \end{equation} 
Since we already know how to
recover $Q$ and $P$ from $g(k,x)$ via \eqref{4.23} and \eqref{4.24}, the same procedure also describes the recovery 
of $Q$ and $P$ from the solution
$\Phi_-(k,x)$ to the Riemann--Hilbert problem.    

In the reflectionless case, i.e. when all the four reflection coefficients are zero, the Riemann--Hilbert problem posed in \eqref{4.2} reduces to
\begin{equation}\label{4.28}
\Phi_+(k,x)=\Phi_-(k,x), \qquad k\in\mathcal L,
\end{equation}
as a result of the fact that the jump $J(k,x)$ given in \eqref{4.7} vanishes. This special case is important because the solution to the Riemann--Hilbert
problem posed in \eqref{4.28} yields soliton solutions to various integrable evolution equations. We refer the reader to \cite{ACTU2025}, when
soliton solutions to various integrable evolution equations, including the Sawada--Kotera equation \cite{SK1974} and a modified form of the bad 
Boussinesq \cite{B1872} equation, are constructed by solving the Riemann--Hilbert problem \eqref{4.28} with an appropriate
input data set consisting of the left transmission coefficient, the bound-state poles of the left transmission coefficient, and the bound-state 
dependency constants with the appropriate time evolution.

\section{The Marchenko integral equation}
\label{section5}

In this section, when the secondary reflection coefficients $M(k)$ and $N(k)$ are both zero, we show how the Riemann--Hilbert problem formulation of
the inverse scattering problem for \eqref{1.1} can be converted to a linear integral equation, which is the analog of the Marchenko integral equation \cite{CS1989,DT1979,F1967,L1987,M2011} arising in
the inverse scattering theory for the full-line Schr\"odinger equation. We derive the Marchenko equation for \eqref{1.1} in the absence of bound states. We
postpone to a future paper the derivation of the Marchenko equation in the presence of bound states. 

By multiplying both sides of \eqref{4.6} with $e^{-kx}$
and subtracting 1 from each side, we obtain 
\begin{equation}\label{5.1}
e^{-kx}\,\Phi_+(k,x)-1=e^{-kx}\,\Phi_-(k,x)-1+e^{-kx}\,J(k,x), \qquad k\in\mathcal L, \quad x\in\mathbb R.
\end{equation}
Let us write \eqref{5.1} as
\begin{equation}\label{5.2}
F(x,s)=G(x,s)+H(x,s), \qquad s\in\mathbb R,
\end{equation}
where we have defined
\begin{equation}\label{5.3}
F(x,s):=e^{-zsx}\Phi_+(zs,x)-1, \qquad s\in\mathbb R,
\end{equation}
\begin{equation}\label{5.4}
G(x,s):=e^{-zsx}\Phi_-(zs,x)-1, \qquad s\in\mathbb R,
\end{equation}
\begin{equation}\label{5.5}
H(x,s):=e^{-zsx}J(zs,x), \qquad s\in\mathbb R,
\end{equation}
by recalling that we have $k=zs$ for $s\in\mathbb R$ when $k\in\mathcal L.$
In the absence of bound states, we already know \cite{ACTU2025,T2024} that for each fixed $x\in\mathbb R,$ the quantity $f(k,x)$ is analytic in $k\in\Omega_1$ and continuous in
$k\in\overline{\Omega_1},$ and it behaves as $e^{kx}[1+O(1/k)]$ as $k\to\infty$ in $\overline{\Omega_1}.$ Similarly, the quantity $g(k,x)$ is analytic in
$k\in\Omega_3$ and continuous in $k\in\overline{\Omega_3},$ and it behaves as $e^{kx}[1+O(1/k)]$ as $k\to\infty$ in $\overline{\Omega_3}.$ In the absence
of bound states, the quantity $m(k,x)$ is analytic $k\in\Omega_2$ and continuous in $k\in\overline{\Omega_2},$ and it has the spectral asymptotics
$e^{kx}[1+O(1/k)]$ as $k\to\infty$ in $\overline{\Omega_2}.$ Similarly, the quantity $n(k,x)$ is analytic $k\in\Omega_4$ and continuous in 
$k\in\overline{\Omega_4},$ and it has the spectral asymptotics $e^{kx}[1+O(1/k)]$ as $k\to\infty$ in $\overline{\Omega_4}.$ In fact, those large $k$-asymptotics
are already indicated in \eqref{2.22}--\eqref{2.25}. Thus, for each fixed
$x\in\mathbb R$ it follows that the quantity $F(x,s)$ is analytic in $s\in\mathbb C^+$ and continuous in $s\in\overline{\mathbb C^+},$ and we have
$F(x,s)=O(1/s)$ as $s\to\infty$ in $s\in\overline{\mathbb C^+}.$ Similarly, for each fixed $x\in\mathbb R$ it follows that the quantity $G(x,s)$ is analytic in
$s\in\mathbb C^-$ and continuous in $s\in\overline{\mathbb C^-},$ and we have $G(x,s)=O(1/s)$ as $s\to\infty$ in $s\in\overline{\mathbb C^-}.$
Consequently, the Fourier transforms in $L^2(\mathbb R)$ with $s\in\mathbb R$ for $F(x,s)$ and $G(x,s)$ exist. From \eqref{5.2}, it follows that the Fourier
transform of $H(x,s)$ also exists. We define the Fourier transforms $\hat F(x,y)$ and $\hat G(x,y)$ by letting 
\begin{equation}\label{5.6}
\hat F(x,y):=\ds\frac{1}{2\pi} \int_{-\infty}^\infty ds\, e^{-isy}F(x,s), \qquad y\in\mathbb R,
\end{equation}
\begin{equation}\label{5.7}
\hat G(x,y):=\ds\frac{1}{2\pi} \int_{-\infty}^\infty ds\, e^{-isy}G(x,s), \qquad y\in\mathbb R.
\end{equation}
From \eqref{5.6} and \eqref{5.7}, respectively, we get the inverse Fourier transforms given by
\begin{equation}\label{5.8}
F(x,s)=\int_{-\infty}^\infty dy\, e^{isy}\hat F(x,y), \qquad s\in\mathbb R,
\end{equation}
\begin{equation}\label{5.9}
G(x,s)=\int_{-\infty}^\infty dy\, e^{isy}\hat G(x,y), \qquad s\in\mathbb R.
\end{equation}
Because of the analyticity of $F(x,s)$ in $s\in\mathbb C^+,$ it follows that
\begin{equation}\label{5.10}
\hat F(x,y)=0, \qquad y<0.
\end{equation}
Similarly, due to the analyticity of $G(x,s)$ in $s\in\mathbb C^-,$ we have 
\begin{equation}\label{5.11}
\hat G(x,y)=0, \qquad y>0.
\end{equation}
Thus, we can write \eqref{5.8} and \eqref{5.9} in their respective equivalent forms as
\begin{equation}\label{5.12}
F(x,s)=\int_0^\infty dy\, e^{isy}\hat F(x,y), \qquad s\in\mathbb R,
\end{equation}
\begin{equation}\label{5.13}
G(x,s)=\int_{-\infty}^0 dy\, e^{-isy}\hat G(x,y), \qquad s\in\mathbb R.
\end{equation}

Next, we are interested in the analysis of $H(x,s)$ appearing in \eqref{5.5} and its Fourier transform. With the help of \eqref{4.6}, \eqref{4.7}, \eqref{5.1}, and 
\eqref{5.2}, we see that
\begin{equation}\label{5.14}
H(x,s)=\begin{cases}
e^{-zsx}\,L(zs)\,T_{\text{\rm{l}}}(z^2s)\,f(z^2s,x), \qquad s>0,\\
\noalign{\medskip}
-\ds\frac{R(zs)\,T_{\text{\rm{r}}}(z^2s)\,g(z^2s,x)}{T_{\text{\rm{r}}}(zs)}, \qquad s<0.
\end{cases}
\end{equation}
Using \eqref{5.3} and \eqref{5.4} in \eqref{5.14}, we obtain
\begin{equation}\label{5.15}
H(x,s)=\begin{cases}
e^{(z^2-z)sx}L(zs)\left[F(x,zs)+1\right], \qquad s>0,\\
\noalign{\medskip}
-\ds\frac{e^{(z^2-z)sx}R(zs)\,T_{\text{\rm{r}}}(z^2s)\left[G(x,zs)+1\right]}{T_{\text{\rm{r}}}(zs)}, \qquad s<0.
\end{cases}
\end{equation}
We introduce the scalar quantity $\rho(s)$ as
\begin{equation}\label{5.16}
\rho(s):=\begin{cases}
\rho_+(s):=L(zs), \qquad s>0,\\
\noalign{\medskip}
\rho_-(s):=-\ds\frac{R(zs)\,T_{\text{\rm{r}}}(z^2s)}{T_{\text{\rm{r}}}(zs)}, \qquad s<0.
\end{cases}
\end{equation}
From \eqref{5.15} and \eqref{5.16}, we have
\begin{equation}\label{5.17}
H(x,s)=\begin{cases}
e^{-\sqrt{3}isx}\rho_+(s)\left[F(x,zs)+1\right], \qquad s>0,\\
\noalign{\medskip}
e^{-\sqrt{3}isx}\rho_-(s)\left[G(x,zs)+1\right], \qquad s<0.
\end{cases}
\end{equation}
In the absence of bound states, let us show that 
\begin{equation}\label{5.18}
F(x,zs)=0, \qquad s<0,
\end{equation}
\begin{equation}\label{5.19}
G(x,zs)=0, \qquad s>0.
\end{equation}
For this, we proceed as follows. Since $F(x,\eta)$ is analytic in $\eta\in\mathbb C^+,$ continuous in $\eta\in\overline{\mathbb C^+},$ and $O(1/\eta)$ 
as $\eta\to\infty$ in $\eta\in\overline{\mathbb C^+},$ we have
\begin{equation}\label{5.20}
\ds\frac{1}{2\pi i}\int_{-\infty}^\infty \,d\eta\,\ds\frac{F(x,\eta)}{\eta-\alpha}=\begin{cases}
F(x,\alpha), \qquad \alpha\in\mathbb C^+,\\
\noalign{\medskip}
0, \qquad \alpha\in\mathbb C^-.
\end{cases}
\end{equation}
We remark that \eqref{5.20} can be established by writing the integral over $\eta\in\mathbb R$ as an integral over the boundary of 
the upper half complex $\eta$-plane and by using Jordan's lemma from the theory of complex variables \cite{R1987}.
Similarly, since $G(x,\eta)$ is analytic in $\eta\in\mathbb C^-,$ continuous in $\eta\in\overline{\mathbb C^-},$ and $O(1/\eta)$ as
$\eta\to\infty$ in $\eta\in\overline{\mathbb C^-},$ it follows that
\begin{equation}\label{5.21}
\ds\frac{1}{2\pi i}\int_{-\infty}^\infty \,d\eta\,\ds\frac{G(x,\eta)}{\eta-\alpha}=\begin{cases}
0, \qquad \alpha\in\mathbb C^+,\\
\noalign{\medskip}
-G(x,\alpha), \qquad \alpha\in\mathbb C^-.
\end{cases}
\end{equation}
One can establish \eqref{5.21} by writing the integral over $\eta\in\mathbb R$ as an integral over the boundary of the
lower half of the complex $\eta$-plane and by using Jordan's lemma.
From \eqref{2.1} it follows that $zs\in \mathbb C^+$ when $s>0$ and that $zs\in \mathbb C^-$ when $s<0.$ Thus, \eqref{5.20} yields \eqref{5.18}
by choosing $\alpha=zs$ in \eqref{5.20} with $s>0.$
Similarly, \eqref{5.21} yields \eqref{5.19} by choosing $\alpha=zs$ in \eqref{5.21} with $s<0.$

Having established \eqref{5.18} and \eqref{5.19}, we proceed as follows. Using \eqref{5.16}, \eqref{5.18}, 
and \eqref{5.19} in \eqref{5.17}, we get
\begin{equation}\label{5.22}
H(x,s)=e^{-\sqrt{3}isx}\rho(s)+\left[F(x,zs)+G(x,zs)\right]e^{-\sqrt{3}isx}\rho(s), \qquad s\in\mathbb R.
\end{equation}  
We use $\hat\rho(y)$ to denote the Fourier transform of $\rho(s)$ defined in \eqref{5.16}, i.e. we let
\begin{equation}\label{5.23}
\hat\rho(y):=\ds\frac{1}{2\pi}\int_{-\infty}^\infty ds\,e^{-isy}\rho(s), \qquad y\in\mathbb R.
\end{equation}  
From \eqref{5.23}, by using the inverse Fourier transform we obtain
\begin{equation*}
\rho(s)=\int_{-\infty}^\infty dy\,e^{isy}\hat\rho(y), \qquad s\in\mathbb R.
\end{equation*}  
From \eqref{5.16} and \eqref{5.23}, we see that
\begin{equation*}
\hat\rho(y)=\hat\rho_+(y)+\hat\rho_-(y), \qquad y\in\mathbb R,
\end{equation*}  
where we have let
\begin{equation}\label{5.26}
\hat\rho_+(y):=\ds\frac{1}{2\pi}\int_0^\infty ds\,e^{-isy}\rho_+(s), \qquad y\in\mathbb R,
\end{equation}  
\begin{equation}\label{5.27}
\hat\rho_-(y):=\ds\frac{1}{2\pi}\int_{-\infty}^0 ds\,e^{-isy}\rho_-(s), \qquad y\in\mathbb R.
\end{equation}  
With the help of \eqref{5.6}, \eqref{5.7}, \eqref{5.22}, \eqref{5.23}, we obtain 
\begin{equation}\label{5.28}
\hat H(x,y)=\hat \rho(\sqrt{3}x+y)+\int_{-\infty}^\infty d\zeta\left[\hat F(x,\zeta)+\hat G(x,\zeta)\right]\hat\rho(\sqrt{3}x-z\zeta +y), \qquad y\in\mathbb R,
\end{equation}  
where we have let
\begin{equation*}
\hat H(x,y):=\ds\frac{1}{2\pi}\int_{-\infty}^\infty d\zeta\, e^{-isy}H(x,s), \qquad y\in\mathbb R.
\end{equation*}
With the help of \eqref{5.6}, \eqref{5.7}, and \eqref{5.28}, we take the Fourier transform of both sides of
\eqref{5.2}, and we obtain
\begin{equation}\label{5.30}
\hat F(x,y)=\hat G(x,y)+\hat \rho(\sqrt{3}x+y)+\int_{-\infty}^\infty d\zeta\left[\hat F(x,\zeta)+\hat G(x,\zeta)\right]\hat\rho(\sqrt{3}x-z\zeta +y), \qquad y\in\mathbb R.
\end{equation} 
Using \eqref{5.10} and \eqref{5.11} in \eqref{5.30}, we obtain
\begin{equation}\label{5.31}
\begin{cases}
\begin{aligned}
\hat F(x,y)=&\hat \rho(\sqrt{3}x+y)\\
                &+\ds\int_0^\infty d\zeta\,\hat F(x,\zeta)\,\hat\rho(\sqrt{3}x-z\zeta +y)
                +\ds\int_{-\infty}^0 d\zeta\,\hat G(x,\zeta)\,\hat\rho(\sqrt{3}x-z\zeta +y), \qquad y>0,
\end{aligned}    \\
\noalign{\medskip}
\begin{aligned}
0=&\hat G(x,y)+\hat \rho(\sqrt{3}x+y)\\
    &+\ds\int_0^\infty d\zeta\,\hat F(x,\zeta)\,\hat\rho(\sqrt{3}x-z\zeta +y)
     +\ds\int_{-\infty}^0 d\zeta\,\hat G(x,\zeta)\,\hat\rho(\sqrt{3}x-z\zeta +y), \qquad y<0.
\end{aligned}                                                  
\end{cases}
\end{equation} 
We would like to simplify the integral terms in \eqref{5.31} by replacing $\hat\rho(\eta)$ 
with $\hat\rho_=(\eta)$ and $\hat\rho_-(\eta)$ when $\eta>0$ and $\eta<0,$ respectively,
where we recall that $\hat\rho_+(y)$ and $\hat\rho_-(y)$ are the quantities defined in \eqref{5.26} and \eqref{5.27}, respectively. 
In other words, we would like to show that
\begin{equation}\label{5.32}
\int_0^\infty d\zeta\,\hat F(x,\zeta)\,\hat\rho(\sqrt{3}x-z\zeta +y)=\int_0^\infty d\zeta\,\hat F(x,\zeta)\,\hat\rho_+(\sqrt{3}x-z\zeta +y), \qquad y\in\mathbb R,
\end{equation} 
\begin{equation}\label{5.33}
\int_{-\infty}^0 d\zeta\,\hat G(x,\zeta)\,\hat\rho(\sqrt{3}x-z\zeta +y)=\int_{-\infty}^0 d\zeta\,\hat G(x,\zeta)\,\hat\rho_-(\sqrt{3}x-z\zeta +y), \qquad y\in\mathbb R.
\end{equation} 
For the proof of \eqref{5.32}
we proceed as follows. With the help of \eqref{5.23}, we write the left-hand side of \eqref{5.32} as 
\begin{equation}\label{5.34}
\int_0^\infty d\zeta\,\hat F(x,\zeta)\,\hat\rho(\sqrt{3}x-z\zeta +y)
=\int_0^\infty d\zeta\,\hat F(x,\zeta)\int_{-\infty}^\infty \ds\frac{ds}{2\pi}\,e^{-is(\sqrt{3}x-z\zeta +y)}\rho(s), \qquad y\in\mathbb R.
\end{equation} 
By changing the order of integration on the right-hand side of \eqref{5.34}, we get  
\begin{equation}\label{5.35}
\int_0^\infty d\zeta\,\hat F(x,\zeta)\,\hat\rho(\sqrt{3}x-z\zeta +y)
=\int_{-\infty}^\infty \ds\frac{ds}{2\pi}\,e^{-is(\sqrt{3}x+y)}\rho(s)\int_0^\infty d\zeta\,\hat F(x,\zeta)\,e^{isz\zeta} , \qquad y\in\mathbb R.
\end{equation} 
Using \eqref{5.12} on the right-hand side of \eqref{5.35}, we obtain 
\begin{equation}\label{5.36}
\int_0^\infty d\zeta\,\hat F(x,\zeta)\,\hat\rho(\sqrt{3}x-z\zeta +y)
=\int_{-\infty}^\infty ds\,e^{-is(\sqrt{3}x+y)}\rho(s)\,F(x,zs), \qquad y\in\mathbb R.
\end{equation} 
From \eqref{5.18}, we know that $F(x,zs)$ vanishes when $s<0.$ Hence, we can write \eqref{5.36} as
\begin{equation}\label{5.37}
\int_0^\infty d\zeta\,\hat F(x,\zeta)\,\hat\rho(\sqrt{3}x-z\zeta +y)
=\int_0^\infty ds\,e^{-is(\sqrt{3}x+y)}\rho(s)\,F(x,zs), \qquad y\in\mathbb R.
\end{equation} 
Using \eqref{5.16} and \eqref{5.26} on the right-hand side of \eqref{5.37}, we get
\begin{equation}\label{5.38}
\int_0^\infty d\zeta\,\hat F(x,\zeta)\,\hat\rho(\sqrt{3}x-z\zeta +y)
=\int_0^\infty ds\,e^{-is(\sqrt{3}x+y)}\rho_+(s)\,F(x,zs), \qquad y\in\mathbb R.
\end{equation}  
Since $F(x,zs)=0$ for $s<0,$ we can write the right-hand side of \eqref{5.38} over the integral on $s\in\mathbb R,$ and hence we get  
\begin{equation}\label{5.39}
\int_0^\infty d\zeta\,\hat F(x,\zeta)\,\hat\rho(\sqrt{3}x-z\zeta +y)
=\int_{-\infty}^\infty ds\,e^{-is(\sqrt{3}x+y)}\rho_+(s)\,F(x,zs), \qquad y\in\mathbb R.
\end{equation}  
Finally, comparing \eqref{5.39} with \eqref{5.36}, we conclude that 
\begin{equation*}
\int_0^\infty d\zeta\,\hat F(x,\zeta)\,\hat\rho(\sqrt{3}x-z\zeta +y)
=\int_0^\infty d\zeta\,\hat F(x,\zeta)\,\hat\rho_+(\sqrt{3}x-z\zeta +y), \qquad y\in\mathbb R,
\end{equation*}  
which coincides with \eqref{5.32}. The proof of \eqref{5.33} is obtained in a similar manner by exploiting the fact that $G(x,zs)=0$ when $s>0,$
which is stated in \eqref{5.19}.
Thus, using \eqref{5.32} and \eqref{5.33} in \eqref{5.31}, we obtain the Marchenko integral equation for \eqref{1.1} as 
\begin{equation}\label{5.41}
\begin{cases}
\begin{aligned}
\hat F(x,y)=&\hat \rho_+(\sqrt{3}x+y)\\
               &+\ds\int_0^\infty d\zeta\,\hat F(x,\zeta)\,\hat\rho_+(\sqrt{3}x-z\zeta +y)
            +\ds\int_{-\infty}^0 d\zeta\,\hat G(x,\zeta)\,\hat\rho_-(\sqrt{3}x-z\zeta +y), \qquad y>0,
\end{aligned}                                                  \\
\noalign{\medskip}
\begin{aligned}
0=&\hat G(x,y)+\hat \rho_-(\sqrt{3}x+y)\\
                    &+\ds\int_0^\infty d\zeta\,\hat F(x,\zeta)\,\hat\rho_+(\sqrt{3}x-z\zeta +y)
               +\ds\int_{-\infty}^0 d\zeta\,\hat G(x,\zeta)\,\hat\rho_-(\sqrt{3}x-z\zeta +y), \qquad y<0.
\end{aligned}                                                  
\end{cases}
\end{equation} 

Having obtained the Marchenko integral equation for \eqref{1.1} given in \eqref{5.41}, we now show how to recover the potentials $Q$ and $P$
from the solution $\hat F(x,y)$ to the Marchenko equation. For this, we proceed as follows. From \eqref{5.12} we get
\begin{equation}\label{5.42}
F(x,s)=\int_0^{\infty} dy\,\hat F(x,y)\,\ds\frac{d}{dy}\left(\ds\frac{e^{isy}}{is}\right).
\end{equation} 
Using integration by parts on the right-hand side of \eqref{5.42}, we obtain
\begin{equation}\label{5.43}
F(x,s)=\hat F(x,y)\,\ds\frac{e^{isy}}{is}\Biggr|_{y=0^+}^{+\infty}-\ds\frac{1}{is}\int_0^{\infty} dy\,\hat F_y(x,y)\,e^{isy},
\end{equation}
where the subscript $y$ denotes the partial derivative with respect to $y.$ From \eqref{5.43} we get
\begin{equation}\label{5.44}
F(x,s)=-\ds\frac{1}{is}\,\hat F(x,0^+)-\ds\frac{1}{is}\int_0^{\infty} dy\,\hat F_y(x,y)\,e^{isy},\qquad s\in\mathbb R.
\end{equation}
A second integration by parts on the right-hand side of \eqref{5.44} yields
\begin{equation}\label{5.45}
F(x,s)=-\ds\frac{1}{is}\,\hat F(x,0^+)-\ds\frac{1}{s^2}\,\hat F_y(x,0^+)+O\left(\ds\frac{1}{s^3}\right),\qquad s\to\pm\infty.
\end{equation}
We recall that $k=zs$ for $s\in\mathbb R$ when $k\in\mathcal L,$ and hence we can write \eqref{5.45} as
\begin{equation}\label{5.46}
e^{-kx}\Phi_+(k,x)-1=-\ds\frac{1}{iz^2k}\,\hat F(x,0^+)-\ds\frac{1}{zk^2}\,\hat F_y(x,0^+)+O\left(\ds\frac{1}{k^3}\right),\qquad k\to\infty
 \text{\rm{ on }}
\mathcal L_1,
\end{equation}
where we have used \eqref{5.3} on the left-hand side and used $z^4=z$ on the right-hand side. In fact, the asymptotics given in \eqref{5.46} holds when $k\in\overline{\Omega_1}.$ 
Comparing \eqref{5.46} with \eqref{4.26}, we obtain
\begin{equation}\label{5.47}
\begin{cases}
t_{\text{\rm{l}}1}+u_1(x)=-\ds\frac{1}{iz^2}\,\hat F(x,0^+),\\
\noalign{\medskip}
t_{\text{\rm{l}}2}+t_{\text{\rm{l}}1}\,u_1(x)+u_2(x)=-\ds\frac{1}{z}\,\hat F_y(x,0^+),
\end{cases}
\end{equation}
where we recall that $t_{\text{\rm{l}}1}$ and $t_{\text{\rm{l}}2}$ are the constants appearing in \eqref{4.25}. By solving the linear algebraic system 
\eqref{5.47} for $u_1(x)$ and $u_2(x),$ we obtain
\begin{equation}\label{5.48}
u_1(x)=-t_{\text{\rm{l}}1}+iz\,\hat F(x,0^+), \qquad x\in\mathbb R,
\end{equation}
\begin{equation}\label{5.49}
u_2(x)=-t_{\text{\rm{l}}2}+t_{\text{\rm{l}}1}^2-iz\,t_{\text{\rm{l}}1}\hat F(x,0^+)-z^2\hat F_y(x,0^+), \qquad x\in\mathbb R.
\end{equation}
Finally using \eqref{5.48} and \eqref{5.49} in \eqref{4.25} and \eqref{4.26} we recover the potentials $Q$ and $P$ from the solution $\hat F(x,y)$
to the Marchenko equation as
\begin{equation*}
Q(x)=-3iz\,\ds\frac{d\hat F(x,0^+)}{dx}, \qquad x\in\mathbb R,
\end{equation*}
\begin{equation*}
P(x)=-3z^2\hat F(x,0^+)\,\ds\frac{d\hat F(x,0^+)}{dx}-3iz\,\ds\frac{d^2\hat F(x,0^+)}{dx^2}+3z^2\,\ds\frac{d\hat F_y(x,0^+)}{dx}, \qquad x\in\mathbb R.
\end{equation*}

Alternatively, we can recover the potentials $Q$ and $P$ from the solution $\hat G(x,y)$ to the Marchenko equation. For this, we proceed as follows. From 
\eqref{5.13} we get
\begin{equation}\label{5.52}
G(x,s)=\int_{-\infty}^0 dy\,\hat G(x,y)\,\ds\frac{d}{dy}\left(\ds\frac{e^{isy}}{is}\right).
\end{equation}
Using integration by parts on the right-hand side of \eqref{5.52}, we obtain
\begin{equation*}
G(x,s)=\hat G(x,y)\,\ds\frac{e^{isy}}{is}\Biggr|_{y=-\infty}^{0^-}-\ds\frac{1}{is}\int_{-\infty}^0 dy\,\hat G_y(x,y)\,e^{isy}, \qquad s\in\mathbb R,
\end{equation*}
or equivalently
\begin{equation}\label{5.54}
G(x,s)=\ds\frac{1}{is}\,\hat G(x,0^-)-\ds\frac{1}{is}\int_{-\infty}^0 dy\,\hat G_y(x,y)\,e^{isy},\qquad s\in\mathbb R.
\end{equation}
Another integration by parts on the right-hand side of \eqref{5.54} yields
\begin{equation}\label{5.55}
G(x,s)=\ds\frac{1}{is}\,\hat G(x,0^-)+\ds\frac{1}{s^2}\,\hat G_y(x,0^-)+O\left(\ds\frac{1}{s^3}\right),\qquad s\to\pm\infty.
\end{equation}
Since $k=zs$ for $s\in\mathbb R$ is equivalent to $k\in\mathcal L,$ from \eqref{5.55} we obtain
\begin{equation}\label{5.56}
e^{-kx}\Phi_-(k,x)-1=\ds\frac{1}{iz^2k}\,\hat G(x,0^-)+\ds\frac{1}{zk^2}\,\hat G_y(x,0^-)+O\left(\ds\frac{1}{k^3}\right),\qquad k\to\infty
 \text{\rm{ on }}
\mathcal L_3,
\end{equation}
where we have used $z^4=z$ on the right-hand side.
In fact, the asymptotic expression given in \eqref{5.56} holds when $k\in\overline{\Omega_3}.$ Comparing \eqref{5.56} with \eqref{4.27}, we obtain
\begin{equation}\label{5.57}
v_1(x)=-iz\,\hat G(x,0^-), \qquad x\in\mathbb R,
\end{equation}
\begin{equation}\label{5.58}
v_2(x)=z^2\,\hat G_y(x,0^-), \qquad x\in\mathbb R.
\end{equation}
Finally, using \eqref{5.57} and \eqref{5.58} in \eqref{4.23} and \eqref{4.24} we recover the potentials $Q$ and $P$ as
\begin{equation*}
Q(x)=3iz\,\ds\frac{d\hat G(x,0^-)}{dx}, \qquad x\in\mathbb R,
\end{equation*}
\begin{equation*}
P(x)=-3z^2\,\hat G(x,0^-)\ds\frac{d\hat G(x,0^-)}{dx}+3iz\,\ds\frac{d^2\hat G(x,0^-)}{dx^2}-3z^2\,\ds\frac{d\hat G_y(x,0^-)}{dx}, \qquad x\in\mathbb R.
\end{equation*}

\end{document}